\documentclass[floatfix,prb,aps,twocolumn,showpacs,preprintnumbers,
amsmath,amssymb]{revtex4}
\usepackage{graphicx}
\usepackage{dcolumn}
\usepackage{exscale}
\begin{document}
\title{Dynamical Properties of the  Anderson Impurity Model
 within a Diagrammatic Pseudoparticle Approach}
\author{S. Kirchner}
\affiliation{Department of Physics \& Astronomy, Rice University, 
 Houston, Texas  77005}
\author{J. Kroha}
\affiliation{Physikalisches Institut, Universit\"at Bonn, 53115 Bonn, Germany}
\author{P. W\"{o}lfle}
\affiliation{Institut f\"ur Theorie der  Kondensierten Materie, 
Universit\"at  Karlsruhe,  76128 Karlsruhe,  Germany}
\begin{abstract}
\noindent
The Anderson model of a twofold spin degenerate impurity level in the 
limit of infinite Coulomb repulsion, $U \rightarrow \infty$, coupled to
one and two degenerate conduction bands or channels, 
is considered in pseudo-particle representation. 
We extend the Conserving T-Matrix Approximation (CTMA),
a general diagrammatic approximation scheme based on a fully renormalized 
computation of two-particle vertex functions in the spin and in the 
charge channel, to the calculation of thermodynamic and spectral 
properties. 
In the single-channel case, the CTMA yields in the Kondo regime  
a temperature independent Pauli  
spin susceptibility for temperatures below the Kondo temperature $T_K$ and 
down to the lowest temperatures considered, reproducing the exact 
spin screening in the Fermi liquid state.
The impurity spectral density appears to remain non-singular down to
the lowest temperatures, in agreement with Fermi liquid behavior. 
However, the unitarity sum rule,
which is crucial for an impurity solver like the CTMA to be applicable
within Dynamical Mean Field Theories for strongly correlated
lattice models, is overestimated at the lowest temperatures.
We argue that this shortcoming may be due to numerical imprecision
and discuss an appropriate scheme for its correction.
In the two-channel case, the spectral density calculated within CTMA 
exhibits qualitatively 
the correct non-Fermi liquid behavior at low temperatures,
i.e. a powerlaw singularity.
\end{abstract}

\maketitle

\section{Introduction}
\label{intro}
Over the past two decades the problem of correlated electrons on a lattice has
 emerged as a central theme of condensed matter theory. With the exception of
 one-dimensional systems,
there are no systematic analytical methods available for solving models 
like the
 Hubbard model. A powerful approximation scheme is the Dynamical 
Mean Field Theory (DMFT),
in which the lattice problem is mapped onto an effective single impurity
 Anderson model
(SIAM), with self-consistently determined properties of the conduction 
electrons\cite{Metzner.89,Georges.96}.
It is a nontrivial task to solve these SIAMs. The properties of real Kondo or 
mixed valence impurities in metals are of interest in their own right,
 with recent
emphasis on non-Fermi liquid behavior in multi-channel
models\cite{Garnier.97,Ott.83,Cox.87,Cox.98}.
Various methods have been
successfully applied to solve these models in certain parameter regimes.
The Bethe ansatz (BA)
method allows to calculate the thermodynamic properties of models with a
flat conduction
electron density of states\cite{Tsvelic.83,Andrei.83}. 
Bosonization methods have been used to obtain, for example, the finite-size
spectrum of one- and two-channel impurities\cite{Zarand.00}.
Conformal field theory is a powerful tool to analyze the low energy
excitations of multi-channel models\cite{Nozieres.80,Ludwig.91,Affleck.92}.
The method of continuous unitary transformations \cite{Wegner.94} has been
successfully applied to the Kondo model in the vicinity of the
Toulouse point \cite{Hofstetter.01}.
These analytical methods are complemented by numerical methods
like Quantum Monte Carlo (QMC) simulations (for not too low temperatures and
moderate U)\cite{Hirsch.86,Gubernatis.87a,Gubernatis.87b}
and Wilson's Numerical Renormalization Group (NRG)
which has been very successful for
not too large degeneracies in the spin or charge
channel\cite{Wilson.75,Costi.94c,Costi.96}.
\newline
The difficulty with quantum systems of
the Anderson impurity type is the strong on-site Hubbard repulsion,
which effectively constrains the quantum dynamics to a Hilbert space
with fixed impurity occupation number and 
makes these problems inaccessible by straight-forward
perturbation theory. 
It is, in particular, difficult to describe the weak
coupling (fluctuating local moment) behavior at high energies and the 
strong coupling fixed point behavior, realized below a strong coupling
energy scale, typically the Kondo temperature $T_K$, by a single
technique. 
In view of possible applications as an ``impurity solver''
within DMFT methods or to quantum impurity and quantum dot systems 
with a complex local spectrum, an accurate method which does not
rely on integrability conditions or on the simplicity of the local or
conduction electron spectrum is highly desirable. 
\newline
For that purpose we had proposed earlier a general diagrammatic 
approximation scheme \cite{Kroha.97}.
The starting point is a pseudoparticle representation of the impurity level,
where the constrained dynamics are built into the very definition of 
the quantum fields \cite{Barnes.76}, and approximations conserving its
internal symmetry are defined by means of Luttinger--Ward functionals. 
The conserving approximation which incorporates
the dominant, local spin and charge fluctuations on the level of 
a fully renormalized calculation of the total 
two-particle vertex (or T-matrix) has been termed the 
Conserving T-Matrix Approximation (CTMA). 
In contrast to previous approximations like the Non-Crossing Approximation 
(NCA) \cite{Kuramoto.83,Hartmann.84,Bickers.87a} and its 
extensions \cite{Anders.94},
the CTMA describes the weak as well as the strong coupling
behavior of the single-channel SIAM correctly on the level of the
pseudoparticle propagators \cite{Kroha.97,Schauerte.00}. 
However, physically observable spectral properties had not 
yet been calculated because of their computational complexity.
\newline
In this article we present CTMA results for the thermodynamic
and spectral properties of the SU($N$)$\times$SU($M$) Anderson
impurity model, $N$ being the local spin degeneracy and $M$ the number
of identical, conserved conduction channels. We will
focus on the single-channel Fermi liquid case ($N=2$, $M=1$),
although results for the two-channel non-Fermi liquid sector of the model
($N=2$, $M=2$) will also be shown.
The spin susceptibility as well as the frequency dependence of the impurity
electron selfenergy indicate that the spin--screened Fermi
liquid ground state of the $N=2$, $M=1$ SIAM
is indeed captured by CTMA. However, the unitarity sum rule of
the spectral density, which is vital for DMFT applications, is
overestimated. A detailed inspection of the impurity electron selfenergy 
shows that this failure seems to originate from an imprecise treatment 
of high-energy processes, either due to numerical inaccuracy or due 
to CTMA neglecting non-singular potential scattering terms, and that
such imprecision influences the low-energy 
behavior via the Kramers-Kronig relation. Based on this analysis
we propose below a phenomenological correction scheme which imposes 
the causality of the impurity selfenergy, 
and which may thus make the CTMA applicable as an
impurity solver for DMFT calculations. This approach amounts to adding 
an appropriate potential scattering term 
to the real part of the impurity selfenergy, 
taken to be a temperature independent constant. 
We will term this scheme the ``effective
potential scattering method''.\\
The paper is organized as follows.
We describe the conserving pseudoparticle technique in
section \ref{technique}, including several physical and technical
justifications of the CTMA. The detailed CTMA self-consistent
equations are given in the Appendix.
The CTMA results for the temperature dependent, static spin susceptibility
and for the impurity spectrum are presented in sections \ref{susc}
and \ref{spectrum}, respectively. Our effective potential scattering 
correction scheme is discussed in detail at the end of section
\ref{spectrum}.
In section \ref{sec:non-FL}, we compare the  CTMA and NCA result for the spectral
density of the two-channel SIAM,
where the ground state is not a Fermi liquid.
We conclude with a discussion of the results in section \ref{conclusion}.
\section{Model and Conserving T-Matrix Approximation}
\label{technique}      
We consider the SU($N$)$\times$SU($M$) SIAM in the limit of infinite 
Coulomb repulsion, implying that the $N$-fold degenerate impurity
level (called d-level here), labeled by spin
$\sigma=-N/2,\ldots, +N/2$, is at most singly occupied.
The empty impurity state is $M$-fold degenerate, labeled          
by $\bar\mu=1,\ldots,M$, and is coupled to a corresponding
degenerate degree of freedom in the conduction band, e.g. z-component
of angular momentum.
In the pseudoparticle representation, the
singly occupied (empty) level is created by fermionic (bosonic) operators 
$f_{\sigma}^{\dagger}$ ($b_{\mu}^{\dagger}$), which satisfy the constraint
$Q\,=\, \sum_{\sigma} f_{\sigma}^{\dagger}f_{\sigma}^{ }\,+\,
\sum_{\mu} b_{\mu}^{\dagger}b_{\mu}^{}\,=\,1$. The physical (or $d$) electron
creation operator on the impurity site is 
$d_{\sigma}^{\dagger}=\sum _{\bar\mu}f_{\sigma}^{\dagger}b_{\bar \mu}$.
The SU($N$)$\times$SU($M$)
Anderson impurity Hamiltonian is then defined by      
\begin{eqnarray}
  \label{eq:hamiltonian}
  H\,&=&\, \sum_{{\bf k} \sigma \mu} \epsilon_{{\bf k}}^{} 
c^{\dagger}_{{\bf k} \sigma \mu}c^{}_{{\bf k} \sigma \mu}\,+\, 
\sum_{\sigma}\epsilon_{d,\sigma}
f_{\sigma}^{\dagger}f_{\sigma}^{ } \nonumber \\
&&\,+\, V \sum_{\sigma \mu} (c^{\dagger}_{0 \sigma \mu}
b_{\bar{\mu}}^{\dagger}f_{\sigma}^{ } + h.c.)\,+\, \lambda Q.
\end{eqnarray}
Here, $c^{\dagger}_{0 \sigma \mu}=\sum_{{\bf k}} c_{{\bf k}
 \sigma \mu}^{\dagger}$
creates a conduction electron at the impurity site $\vec{R} =0$ and 
$\epsilon_{d,\sigma}=\epsilon_{d}+\sigma g \mu_B B$ is the
impurity level in a magnetic field $B$, with $\mu_B$ and $g$ the 
Bohr magneton and the Land\'e factor, respectively. The operator $b_{\bar{\mu}}$
transforms according to the conjugate representation  of SU($M$).
We denote the density of states (DOS) of the conduction electrons at the 
Fermi energy $\epsilon _F$ by $N(0)$ and assume it to be structureless. 
(All numerical results were obtained for a Gaussian DOS. Note,
however, that our method works for
arbitrarily structured DOS.) Instead of the hybridization
$V$, we will frequently use $\Gamma=\pi N(0) V^2$ as a parameter of
the model. 
In the Kondo regime, $\Gamma \ll \epsilon _d$ 
where the low energy excitations resemble those of
the Kondo model, the above Hamiltonian has a dynamically generated
strong coupling scale, the Kondo temperature $T_K$, where
perturbation theory breaks down,
 \begin{equation}
   \label{eq:TK}
   T_K\,=\,D \Big(\frac{N \Gamma}{\pi D} \Big)^{(M/N)}\,
   \exp\Big(-\frac{\pi |\epsilon_d|}{N\Gamma}\Big),
 \end{equation}
where $2D$ is the bandwidth of the conduction electron DOS. 
The Hamiltonian  (\ref{eq:hamiltonian}) possesses a U(1) gauge
symmetry with respect to simultaneous transformations of
$f_{\sigma}^{}$ and $b^{}_{\bar\mu}$ related to the conserved
charge $Q$, $\lambda$ being the local gauge field.
The exact projection of the dynamics onto the Hilbert subspace $Q=1$
is accomplished by taking a gauge with a time independent $\lambda$,
re-defining the zero of the energy scale as
$\omega\to\omega + \lambda$, and letting $\lambda\to\infty$ in all
expressions; see Ref. [\onlinecite{Kirchner.PhD}] and 
the appendix of Ref. [\onlinecite{Costi.96}]
for details of the projection technique.
The charge conservation in conjunction with the 
constraint $Q=1$ implies an orthogonality catastrophe between the
$Q=1$ initial and the $Q=0$ final states, and leads to infrared 
threshold power-law behavior of the pseudoparticle Green's functions
$G_{f,b}(\omega) \propto \omega^{-\alpha_{f,b}}$. 
In the Fermi liquid case, $M\leq N-1$, the exponents $\alpha_{f,b}$
are closely related to the average impurity
occupation number $n_d$ via the Friedel sum rule 
\cite{Menge.88,Costi.94a,Costi.94b},
\begin{equation}
\alpha_f = (2n_d^{\phantom{2}} - n_d^2)/N \qquad 
\alpha_b = 1-n_d^2/N
\label{eq:FLexponents}
\end{equation}
Approximation schemes for calculating $G_{f,b}$ which violate the 
gauge symmetry would, hence, violate the orthogonality of initial and
final states, and should be expected to give incorrect results for physical 
quantities, even though certain aspects of the Fermi liquid fixed 
point can be described by symmetry-breaking approximations 
\cite{Read.83,Logan.98,Logan.00,Logan.02,Glossop.02,Glossop.03a,Glossop.03b}. 
Therefore, we take great care to
preserve the gauge symmetry. It can be reconciled with the time
independent choice of the gauge field $\lambda$ necessary for the
$Q=1$ projection by employing
a conserving approximation, derived from a Luttinger-Ward 
generating functional $\Phi$ \cite{Baym.61,Baym.62}.
The local selfenergies $\Sigma_{\alpha}$, $\alpha=f,b,c$, defined by
\begin{eqnarray}
  G_{f\sigma}^{-1}(\omega)&=& \omega-\epsilon_{d\sigma}
                            -\lambda-\Sigma_{f}(\omega) 
\label{eq:Gf}\\
  G_{b}^{-1}(\omega)&=& \omega-\lambda-\Sigma_{b}(\omega) 
\label{eq:Gb}\\
  G_{c\sigma}^{-1}(\omega)&=& G_{c0\sigma}^{-1}(\omega)-
  \Sigma_{c\sigma}(\omega)\ ,
\label{eq:Gc}
\end{eqnarray}
where $G_{c0\sigma}^{-1}(\omega)=\sum_{\vec{k} }(\omega-\epsilon_{k})^{-1}$, 
are generated by functional differentiation of $\Phi$ with respect to
the  {\it self-consistently} renormalized Green`s functions 
$\Sigma_{\alpha}(\omega)=\delta\, \Phi/\delta\, G_{\alpha}(\omega)$. 
As a result, for any given approximation to $\Phi$ a set of
self-consistent nonlinear integral equations for $\Sigma_{f}$ 
and $\Sigma_{b}$ is obtained, which, in general,
cannot be solved analytically but is amenable to numerical solution. 
The central task is then to find the correct generating functional
which captures the essential physics of the problem at hand.
\newline
{\it Non-Crossing Approximation (NCA)}. The NCA is often used
for its computational simplicity to obtain a rough description, and
it even captures the universal behavior inherent to Kondo-type
problems\cite{Bickers.87b}. However, the NCA recovers the correct
Kondo scale $T_K$ only because of a fortunate compensation of the
neglect of spin-flipping logarithmic terms and an incorrect
logarithmic resummation of potential scattering terms
\cite{Kirchner.02,Lebanon.01}. Below $T_K$
it develops spurious infrared singularities in physical quantities.
The NCA breaks down in a
magnetic field $B$ even in the weak coupling regime
($T>T_K$ and/or $B>T_K$), producing, in addition to the
two Zeeman-split Kondo peaks a third, field independent resonance
in the impurity spectral density at the Fermi energy.
It seems that this spurious behavior originates from the
incorrect treatment of the potential scattering in NCA mentioned
above. On the level of auxiliary particles, the NCA does not
give the correct FL threshold exponents Eq.\ (\ref{eq:FLexponents}),
but instead $\alpha_f^{NCA}=M/(N+M)$, $\alpha_b^{NCA} = N/(N+M)$.
It can be shown by power counting arguments, using 
$\alpha_f^{NCA} + \alpha_b^{NCA} = 1$, that any
self-consistent calculation involving only a {\it finite} number of
skeleton selfenergy diagrams just reproduces the incorrect NCA
exponents. 
\newline
{\it Conserving T-Matrix Approximation (CTMA)}.
Hence, selfenergies and two-particle vertex functions must be
comprised of an {\it infinite} class of skeleton diagrams in order to
describe the Fermi liquid fixed point. Since the latter is a
consequence of the singlet formation between conduction electron 
and impurity spin, it is natural to assume that 
higher than two-particle correlations need not be considered 
in the single-channel case.
The total vertex functions of conduction electrons ($c$) and 
local degrees of freedom (pseudofermions $f$, slave bosons $b$) 
are then two-particle 
T-matrices which are obtained from an infinite summation of
irreducible parts via Bethe--Salpeter equations. 
\begin{figure}[h!]
\centerline{\includegraphics[width=1\linewidth]{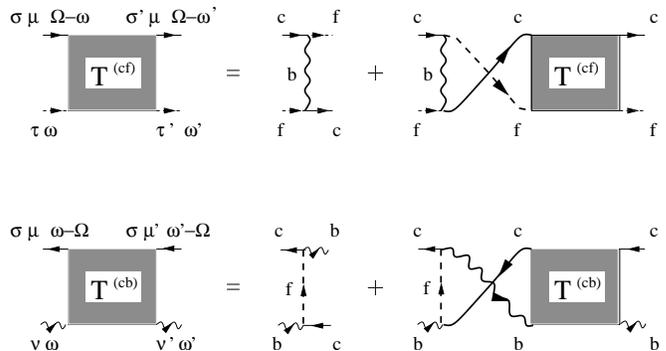}}
\vskip 4pt
\caption{Diagrammatic representation of the
  Bethe-Salpeter equations for the vertex functions
  $T^{(cf)}$ and $T^{(cb)}$ defining the CTMA. Dashed,   
  wiggly, and solid lines represent here and in the following 
  the renormalized pseudofermion,
  slave boson, and the local conduction electron propagators,
  respectively. The external lines are for clarity only and 
  are not part of the vertices. 
}
\label{FIG1}
\end{figure}
\begin{figure}[h!]
\centerline{\includegraphics[width=1\linewidth]{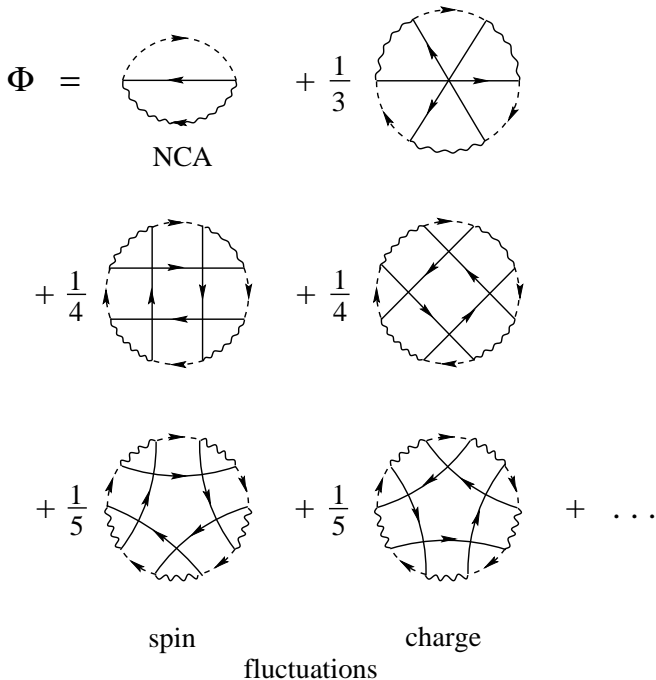}}
\vskip 4pt
\caption{Generating functional of the CTMA. The first diagram
  corresponds to the NCA generating functional. 
  Each diagram contains exactly one closed auxiliary particle loop;
  diagrams with more than one such loop vanish after projection onto $Q=1$.
  The class of CTMA diagrams beyond NCA is uniquely defined 
  by the condition that each conduction electron line spans
  exactly two bare three-point hybridization vertices, as shown.    
  The diagrams of the first and of the second column generate, by 
  second functional derivative, $\delta ^2 \Phi/\delta G_c \delta G_f$ 
  and $\delta ^2 \Phi/\delta G_c \delta G_b$, the spin and the
  charge fluctuation T-matrices, $T^{(cf)}$, $T^{(cb)}$,
  respectively (Fig.\ \ref{FIG1}). }  
\label{FIG2}
\end{figure}
We use the smallness of the
parameter V\,N(0)$ \ll 1$ to select the leading diagrams of the
irreducible parts. This results in the ladder approximation for the
total two-particle vertices shown in Fig.\ \ref{FIG1}. The 
Luttinger--Ward functional that generates by second functional
differentiation the vertex functions of Fig. \ref{FIG1} is constructed 
by connecting the entry and exit points by Green's function lines
and is shown diagrammatically in Fig. \ref{FIG2}. The diagram
containing two (renormalized) boson lines is not a skeleton, is
already contained in the first (NCA) diagram via self-consistency, and,
hence, is omitted. The conserving approximation obtained in this way
has been called Conserving T-Matrix Approximation (CTMA).
The self-consistent equations for the vertex functions and
selfenergies to be solved are given explicitly in Appendix \ref{sec:CTMAequations}.
Note that the $f-c$ and the $b-c$ vertices in Figs. \ref{FIG1} and 
\ref{FIG2} describe spin and 
charge fluctuations, respectively. Therefore the CTMA should
provide a good approximation not only in the Kondo, but also
in the mixed valence and empty impurity regimes. 
\newline
On a more formal level, the CTMA can be justified both near the
weak and near the strong coupling fixed points. Expanding, in the
weak coupling regime, the CTMA in terms of {\it bare}, projected
Green's functions ($B=0$), 
$
  G_{f\sigma}^{0}(\omega)=1/(\omega \pm i0),~  
  G_{b}^{0}(\omega)=1/(\omega+\epsilon_d \pm i0)
$,
it is seen that the CTMA c-f vertex is exact up to leading logarithmic 
order, as seen in Fig. \ref{FIG3}. Therefore, CTMA does 
incorporate the correct renormalization group (RG) flow in the weak
coupling region \cite{Kirchner.02}. In particular, logarithmic
potential scattering terms, present in each one of the diagrams of
Fig. \ref{FIG3}, cancel correctly when the two diagrams are added.
Therefore, we expect that the CTMA correctly describes the Zeeman splitting
of the Kondo resonance even in a large magnetic field, in contrast to
the NCA, where only the first diagram of Fig. \ref{FIG3} is included,
and its logarithmic potential scattering part leads to a spurious
third peak at the the Fermi energy $\omega = 0$ (see above).
\newline
By virtue of the self-consistent inclusion of selfenergy diagrams
in the propagators of Fig.\ \ref{FIG1} the CTMA vertex not only includes
ladder but also parquet diagrams. Moreover, at any given order  
of {\it self-consistent} perturbation theory, the CTMA includes all
diagrams with the maximum number of local spin and charge fluctuation
processes in the sense of principal diagrams. Since these are expected to be
responsible for the formation of the spin-singlet in the Kondo or mixed
valence regime, CTMA may be expected to capture the physics of the
strong coupling fixed point of the single--channel SIAM as well.
This will be shown by numerical evaluation in the following sections.
\begin{figure}[h!]
\centerline{\includegraphics[width=1\linewidth]{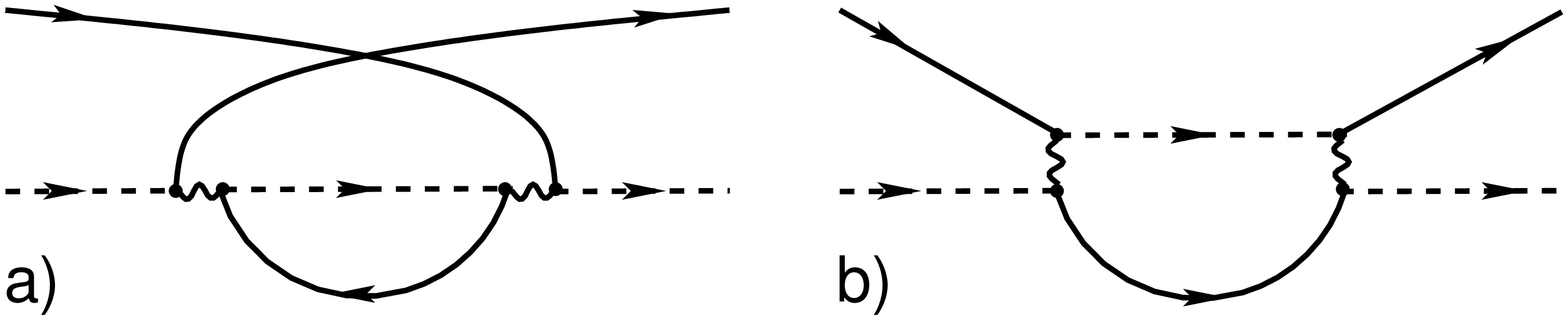}}
\vskip 4pt
\caption{Diagrams containing the leading logarithmic contributions to
the c-f vertex function. In contrast to Figs. \ref{FIG1}, \ref{FIG2},
the lines represent the {\it bare} Green's functions here. 
a) Contribution from the NCA generating functional, b) 
additional contribution from the CTMA. Spurious logarithmic potential
scattering terms cancel only when both terms are added. 
The c-b vertex has no logarithmic terms and,
hence, does not flow under perturbative RG.
}
\label{FIG3}
\end{figure}
\section{Spin Susceptibility}
\label{susc}
\begin{figure}[h!]
\centerline{\includegraphics[width=1\linewidth]{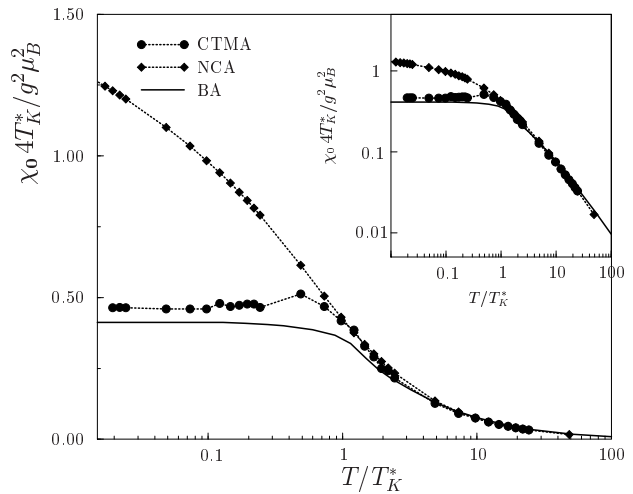}}
\caption{The static magnetic susceptibility as a function of
temperature. The temperature and the susceptibility scale are
plotted in units of the conventional Kondo temperature $T_K^{\star}$
Eq.\ (\ref{eq:TKW}),
that leads to the value $W=0.4128$ for the Wilson number, see text. 
The CTMA curve agrees well
with the exact BA result in both, the high temperature and in the
strong coupling regimes, with only a small spurious intermediate
maximum in the crossover region.}
\label{FIG4}
\end{figure}

In this section we report the 
CTMA result for the static impurity susceptibility
\begin{equation}
\chi_i(T) = \left. \frac{dM}{dB}\right|_{B=0}
\label{eq:susc1}
\end{equation}
of the $N=2$, $M=1$ SIAM in the Kondo regime and compare with NCA and 
Bethe ansatz results. Throughout the paper, the evaluations in this
regime will be done for the typical parameter set $\epsilon _d /D = -0.81$,
$\Gamma /D = 0.2$. In Eq.\ (\ref{eq:susc1}), yielding a low-temperature
impurity occupation number of $n_{d\sigma}\simeq 0.47$ and 
$T_K=4.16 \times 10^{-4}D$. Other parameter sets in the Kondo regime
give similar results. 
$M=g\mu_B\sum_{\sigma}\sigma n_{\sigma}$ is the impurity magnetization,
\begin{equation}
n_{d\sigma} = \lim_{\lambda \to \infty}
\frac{\int d\omega e^{-\beta\omega}\; \mbox{Im} G_{f\sigma}(\omega-i0)}
{\int d\omega e^{-\beta\omega}\; \mbox{Im}[\sum_{\sigma} 
G_{f\sigma}(\omega-i0)+ G_{b}(\omega-i0)]}
\label{eq:nsigma}
\end{equation}
is the impurity occupation number with spin $\sigma$ projected onto $Q=1$
\cite{Kirchner.PhD,Costi.96} ($\beta = 1/k_BT$),
and the magnetic field $B$ couples only to the impurity spin 
(Eq.\ (\ref{eq:hamiltonian})).
The expression (\ref{eq:susc1}) is equivalent to the $\omega = 0$ limit
of the causal dynamical linear response susceptibility 
\begin{equation}
\chi_i(T,\omega=0) 
= -i \int dt \Theta(t) \langle [\hat M(t), \hat M(0)] \rangle \ .
\label{eq:susc2}
\end{equation}
This is readily shown from Eqs. (\ref{eq:susc1}), (\ref{eq:nsigma}),
employing 
$d/dB = \sum _{\sigma} \left( dG_{f\sigma}/dB \right) \delta /
\delta G_{f\sigma}$ and 
\begin{equation}
\frac{dG_{f\sigma}}{dB} = G_{f\sigma}^2 
\left[ \sigma g\mu_B + \sum _{\sigma\sigma '}
\frac{\delta \Sigma _{f\sigma}}{\delta G_{f\sigma '}}\; 
\frac{dG_{f\sigma '}}{dB} \right] \ ,
\label{eq:bethesalpeter}
\end{equation}
which follows from the definition of $G_{f\sigma}$,
\begin{equation}
\frac{\delta \Sigma _{f\sigma}}{\delta G_{f\sigma '}} = \gamma _{\sigma,\sigma '}
\label{eq:WardId}
\end{equation}
is the irreducible four-point pseudofermion vertex. Any conserving
approximation by construction fulfills the equivalence of 
Eqs.\ (\ref{eq:susc1}) and (\ref{eq:susc2}), since it respects the 
Ward identity (\ref{eq:WardId}). We choose to use Eq.\ (\ref{eq:susc1}),
because it is computationally less demanding than the correlation function
Eq.\ (\ref{eq:susc2}).
\newline
Note that additional terms $\chi _b(T)$ and $\chi_{ib}(T)$ arise, if
$B$ couples also to the conduction electron spin. $\chi _b(T)$ is the
constant Pauli susceptibility of the conduction band  and
$\chi_{ib}(T)$ a mixing term correlating the impurity and the conduction
electron magnetization. Since the latter is, for a flat DOS, due to
the electronic polarization at the bottom of the band, $\chi_{ib}(T)$
is usually negligible for $T\ll D$. 
\newline
$\chi_{i}(T)$ is of principal interest as an indicator of whether CTMA
captures the spin singlet Fermi liquid ground state of the 
single-channel SIAM. The result is shown in Fig.\ \ref{FIG4}. While at
exponentially high temperature, $\ln (T/T_K)\gg 1$, 
$\chi_i(T) = (1/4) g^2_{}\mu _B^2/T$, typical for a 
free, fluctuating local moment, $\chi_i(T)$ shows $T$-independent 
Pauli behavior for $T\lesssim 0.1\; T_K$ and down to the lowest $T$
considered, characteristic for the completely spin-screened 
Fermi liquid state. By contrast, the low--$T$ behavior of the NCA
is a power law, $\chi_i^{NCA}(T)- \chi_i^{NCA}(0) \propto - T^{1/3}$.
At $T=0$, $\chi_i (T)$ acquires the value
\begin{equation}
\chi _i(0) = \frac{(g\mu_B)^2}{4 T_L}\ ,
\label{eq:chi0}
\end{equation}    
which defines the universal low-temperature scale, $T_L$, of the Kondo or
Anderson model, related to $T_K$ by the Wilson number $W=T_K/T_L$.

The comparison of the CTMA result with exact methods like NRG 
\cite{Wilson.75} or BA \cite{Tsvelic.83,Andrei.83,Andrei.84}
can be made quantitative. The dimensionless quantity 
$\overline\chi_i(T/T_K)=\chi _i(T)/(g^2_{}\mu_B^2/T_K)$ is known to 
be a universal function of $T/T_K$ (i.e. independent of the 
microscopic parameters of the Hamiltonian), with $\overline\chi_i(0)=W$. 
Self-consistent approaches
like NCA or CTMA reproduce this universality, since they include
a resummation of the logarithmic terms of perturbation theory 
\cite{Rosch.01}. In comparing the CTMA and the BA results one must, 
however, observe, 
that the breakdown scale of perturbation theory, $T_K$, 
depends on its precise definition. Therefore, care must be taken 
that the same definition of $T_K$ is used for both, the CTMA and 
the exact method.
In Wilson's original work on the Kondo model \cite{Wilson.75}, 
a Kondo temperature $T_K^{\star}$ was defined such that
in the high temperature expansion of $\chi_i(T)$ all terms of 
$O(\ln (T/T_K^{\star})^{-2})$ cancel each other. Rasul and Hewson 
\cite{Hewson.83,Rasul.84a,Rasul.84b} used the same criterion for the
SIAM and found for the Kondo temperature,
\begin{equation}
  \label{eq:TKW}
T_K^{\star}=\frac{1}{2 \pi} \exp\left(1+C-\frac{1}{2N}\right) 
\sqrt{\frac{D}{|\epsilon_d|}}\, T_K
\end{equation}
where $C=0.5772157$ is Euler's constant. 
With this definition, the universal Wilson number was found to be
W=0.4128. Using the same definition, we find within CTMA (Fig.\ \ref{FIG4}), 
\begin{equation}
W^{(CTMA)}=T_K^{\star} \chi_i(0)/\mu_B^2 = 0.462\ .
\end{equation}
In the BA method, $T_K$ is defined in a somewhat different way
due to a different cutoff scheme, resulting in a Wilson number 
$W^{BA}= \exp (C+1/4)/\pi ^{3/2} \approx 0.4107$
\cite{Andrei.83,Andrei.84} (for $N=2$). Therefore, in the BA curve for 
$\overline\chi _i(T/T_K)$  we rescale $T_K$ such that 
$\overline\chi_i(0)$ obtains Wilson's $T=0$ value $W=0.4128$
(Fig.\ \ref{FIG4}).  
As seen from the figure, the CTMA result for the static susceptibility
is in strikingly good quantitative agreement with the BA result 
not only for $T>T_K^{\star}$, but also in the strong coupling region, 
$T\lesssim 0.1\; T_K^{\star}$. This shows that the CTMA describes
the low-energy excitations around the Fermi liquid fixed point 
even quantitatively correctly at least in a thermodynamic quantity
like the magnetic susceptibility. From the general properties of 
conserving approximations, one may expect the same to be
true for dynamical quantities as well. This will be investigated in
the next section.

\section{Dynamical Properties}
\label{spectrum}
The Pauli behavior of the impurity susceptibility calculated
in the previous section shows that the spin structure of the
low-$T$ excitations captured by the CTMA free energy functional
is such that it describes the complete screening of the
impurity spin by the conduction electrons correctly even with
good quantitative precision. It is the resulting absence of
spin flip scattering at energies below $T_K$ that is responsible
for the Fermi liquid behavior near the strong coupling fixed
point of the single-channel Kondo or Anderson model. Therefore,
one may conjecture, that the CTMA may capture the Fermi liquid
nature of the low-$T$ excitations as well, with well defined
quasiparticles which should become visible in dynamical
quantities like the impurity spectral density and the selfenergy.
This expectation is further supported by the fact that the CTMA
does indeed reproduce the correct Fermi liquid threshold
exponents on the level of the auxiliary particle dynamics
\cite{Kroha.97}.

The quantities of prime interest, e.g., for DMFT are the
impurity electron Green's function,
\begin{equation} 
G_{d\sigma}(\omega \pm i0) = \frac{1}
{\omega -\epsilon_d \pm i\Gamma - 
\Sigma _{d\sigma}(\omega\pm i0)}\ , 
\label{eq:DefGd}
\end{equation}
and the interaction part of the selfenergy, 
$\Sigma _{d\sigma}(\omega)$, in the spin--screened case, $N=2$, $M=1$.
The Fermi liquid theory implies certain exact low-energy properties, 
which the CTMA results must be compared to, 
namely the unitarity limit for the impurity spectral
density at the Fermi level $\tilde{A}_{d\sigma}(0)$,  
the half width $\tilde\Gamma$ and position 
$\tilde\epsilon_d$ of the Kondo resonance, as well
as the low-energy behavior of $\mbox{Im}\Sigma _{d\sigma}(\omega)$,
\begin{eqnarray}
\tilde{A}_{d\sigma}(0) &=& \frac{\sin ^2(\pi n_{d\sigma})}{\pi\Gamma}, 
\qquad T=0  \label{eq:Ad0}\\
\tilde\Gamma &=& \frac{4}{\pi W}\; \sin^2 (\pi n_{d\sigma})\;
T_K^{\star} \label{eq:Gtilde}\\
\tilde\epsilon _d &=& \frac{2}{\pi W}\;  \sin (2\pi n_{d\sigma})\; 
T_K^{\star} \label{eq:etilde}\\
\mbox{Im} \Sigma _{d\sigma} (\omega-i0) &=& a\; \Gamma\; 
\frac{\omega^2 + (\pi T)^2}{T_K^2} \label{eq:ImSigma}\\
a &=& \frac{\pi^4 W^2}{8 \mbox{e}^{3/2+2C}} \; 
\frac{(R-1)^2}{\sin^2(\pi n_{d\sigma})}\; 
\frac{|\epsilon _d|}{D} \ , \label{eq:prefactor}
\end{eqnarray}
where $R=2$ is the Wilson ratio.
The proof of these relations is compiled in Appendix \ref{sec:FLrelations}.

\subsection{CTMA solution and Fermi liquid behavior}
For the computation of $G_{d\sigma}(\omega)$ within the CTMA, observe that it 
is related to the single-particle conduction electron
T-matrix, $t_{c\sigma} (\omega) = V^2 G_{d\sigma}(\omega)$, where,
after projection onto $Q=1$ only diagrams with a single 
pseudoparticle loop, i.e. irreducible diagrams, remain. In the 
conserving scheme it is, therefore, constructed as
\begin{equation}
G_{d\sigma}(\omega) = \frac{1}{V^2}\; \lim _{\lambda\to\infty}
\frac{1}{Q(\lambda)}\;
\frac{\delta \Phi}{\delta G_{c\sigma}(\omega)}\ .
\label{eq:Gd}
\end{equation} 
The corresponding CTMA diagrams are shown in Fig.\ \ref{FIG5}, and the 
details about their evaluation are given in Appendix \ref{sec:CTMAequations}.
\begin{figure}[b!]
\centerline{\includegraphics[width=1\linewidth]{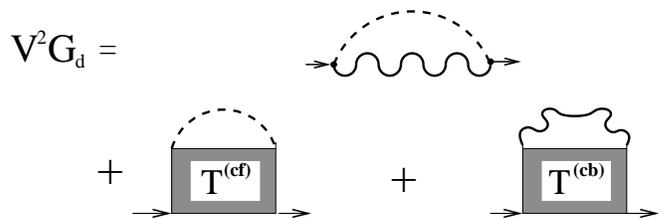}}
\vskip 4pt
\caption{Diagrams defining the d-electron Green's function
within CTMA. For details of their evaluation see Appendix 
\ref{sec:CTMAequations}.
 }
\label{FIG5}
\end{figure}

It is worth noting that, by definition, the impurity electron propagator
is equivalent to the $f-b$--``particle-hole'' correlation function,
$G_{d\sigma}(t)= -i \langle \hat T \left\{ 
b^{\dagger}(t)f_{\sigma}^{\phantom{\dagger}}(t)
f_{\sigma}^{\dagger}(0)b^{\phantom{\dagger}}(0) \right\}\rangle $.
This might seem to offer another possibility of calculating 
$G_{d\sigma}$ using the irreducible $f-b$ vertex
$\Gamma _{fb} = \delta^2 \Phi /\delta G_{f\sigma}\delta G_b$.
However, any diagram of the $f-b$ {\it particle-hole} correlator
constructed in this way contains necessarily two pseudoparticle loops, 
and, hence, vanishes by projection. Therefore, in CTMA
the $f-b$ correlation function is comprised of the 
(NCA-like) $f-b$ bubble diagram only, which is clearly not sufficient to
recover the Fermi liquid strong coupling fixed point.
We note in passing that the non-trivial contributions to $\Gamma _{fb}$
comprising the full $G_d$ are generated from free energy diagrams which 
contain more than one pseudoparticle loop, but are not contained in CTMA. 
As a conclusion, in CTMA $G_{d\sigma}(\omega)$ must be 
calculated unambiguously using Eq.\ (\ref{eq:Gd}). 
\begin{figure}[h!]
\centerline{\includegraphics[width=1\linewidth]{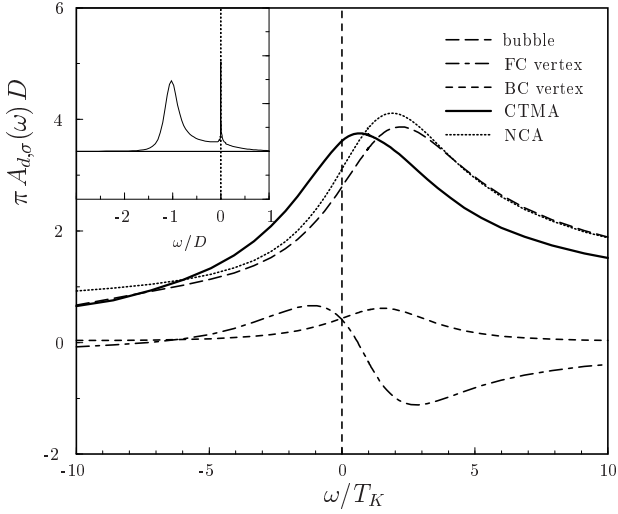}}
\caption{CTMA and NCA results 
for the local spectral density $A_{d\sigma}(\omega)$
for $\epsilon_d/D = -0.81$, $\Gamma /D= 0.2$ at $T=T_K=6.16\times 10^{-4}D$.
The decomposition of the CTMA result into $f-b$ bubble and vertex
corrections arising from  $T^{(cf)}$ (fc vertex)
and $T^{(cb)}$ (bc vertex) is also shown. The NCA shows an incorrectly
large shift of the Kondo peak of $O(T_K)$ towards positive frequencies,
which is corrected by CTMA; see the text.}
\label{FIG6}
\end{figure}
\begin{figure}[h!]
\centerline{\includegraphics[width=1\linewidth]{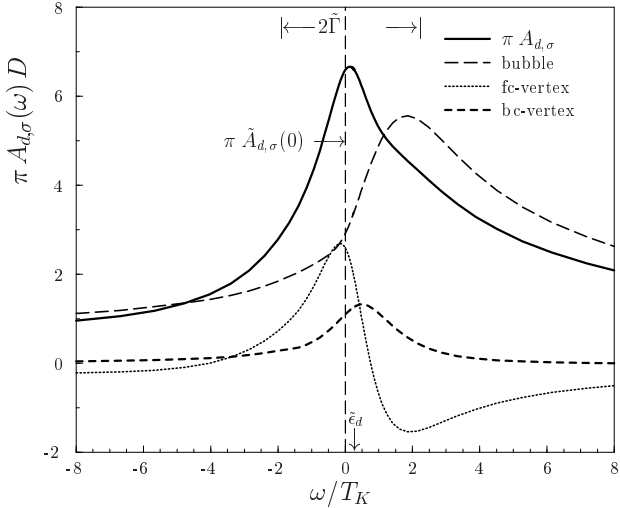}}
\caption{CTMA impurity spectral function and decomposition into 
$f-b$ bubble and vertex contributions for the same parameters as
in Fig.\ \ref{FIG6}, but at $T=0.01T_K$. The exact values for 
$A_{d\sigma}(0)$, $\tilde\Gamma$, $\tilde\epsilon _d$, Eq.\ 
(\ref{eq:Ad0}), (\ref{eq:Gtilde}), (\ref{eq:etilde}), 
are displayed for comparison.}
\label{FIG7}
\end{figure}

\begin{figure}[h!]
\centerline{\includegraphics[width=1\linewidth]{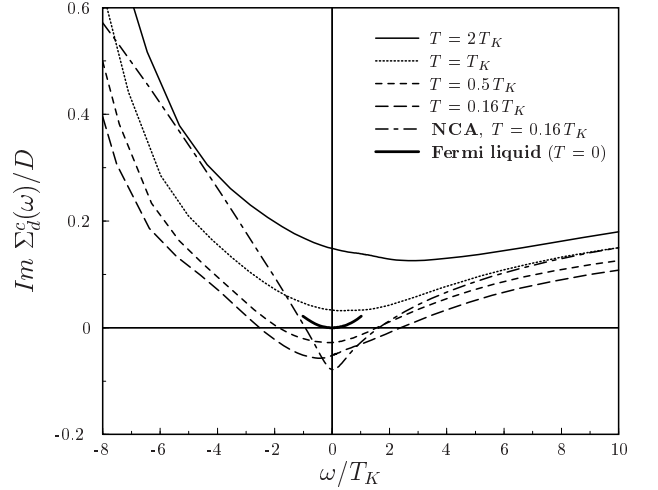}}
\caption{Correlation part of the physical $d-$electron
selfenergy $\mathop{\rm Im}\Sigma _{d\sigma}(\omega)$  calculated in 
CTMA for various temperatures and $\epsilon_d/D=-0.81$, $\Gamma/D=0.2$. 
For the lowest temperature shown, $T=0.16\, T_K$, the NCA result 
(dot--dashed line) is shown for comparison, showing the cusp-like
infrared singularity typical for NCA. 
The bold solid line centered at the origin represents the $T=0$ 
behavior expected for our parameter set based on Fermi liquid theory,
Eqs.\ (\ref{eq:ImSigma}), (\ref{eq:prefactor}).}
\label{FIG8}
\end{figure}

\begin{figure}[t!]
\centerline{\includegraphics[width=1\linewidth]{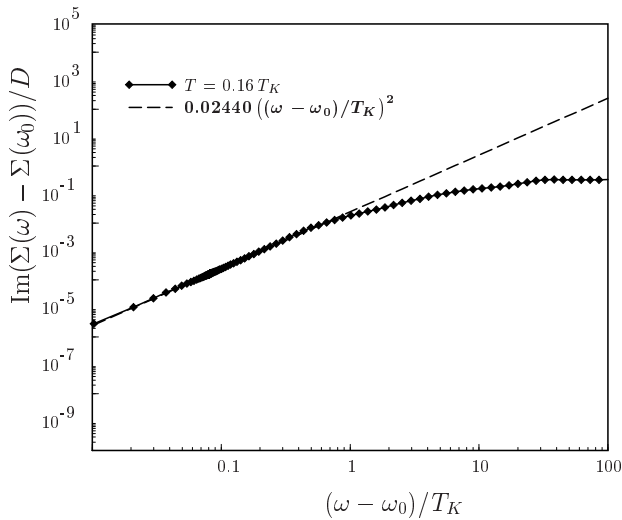}}
\caption{Log$_{10}$-log$_{10}$ plot of 
$\mathop{\rm Im} \Sigma_{d\sigma}(\omega  -i0) -
\mathop{\rm Im} \Sigma_{d\sigma}(\omega _0 -i0)$
versus frequency, $(\omega -\omega _0)/T_K$, for $T=0.16\ T_K$. $\omega _0$ is the 
position of the minimum of $\mathop{\rm Im} \Sigma_{d\sigma} 
(\omega -i0)$ in Fig.\ \ref{FIG8}. The dashed line is the
fit to the low-frequency quadratic behavior of 
$\mathop{\rm Im} \Sigma_{d\sigma}(\omega  -i0)$ and represents the
function \mbox{$y = 0.02440 (\omega - \omega _0)^2 /T_K^2$}, see the text.  
\label{FIG9}}
\end{figure}

\begin{figure}[h!]
\vskip 10pt
\centerline{\includegraphics[width=1\linewidth]{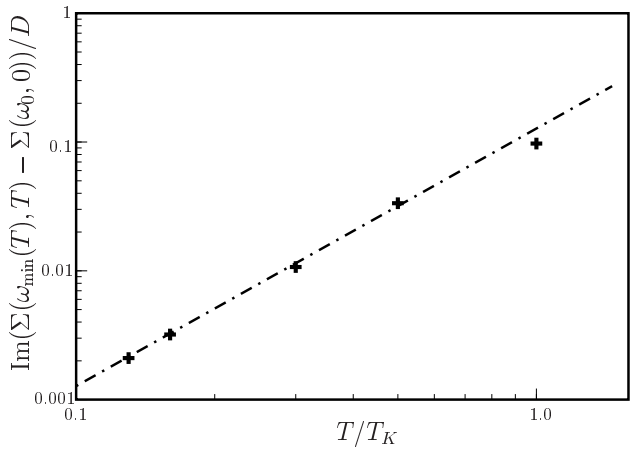}}
\caption{Log$_{10}$-log$_{10}$ plot of the minimum value of the
 imaginary part of
the selfenergy,
$\mathop{\rm Im} \Sigma _{d\sigma}(\omega = \omega _{min}(T), T) 
-\mathop{\rm Im} \Sigma _{d\sigma}(\omega = \omega _0, T = 0.01T_K)$
versus $T/T_K$. $\omega _{min}(T)$ is the position of the minimum
for a given temperature $T$, see Fig.\ \ref{FIG8}. Quadratic in $T$
behavior is clearly visible for  $T \lesssim T_K$. 
The dashed line is given by $y=0.013\,D\, \big(\pi\,T/T_K
\big)^2$. 
\label{FIG10}}
\end{figure}

The CTMA results for the d--electron spectral function,
$A_{d\sigma}(\omega) = \mbox{Im} G_{d\sigma} (\omega -i0)/\pi$ are
shown in Figs.\ \ref{FIG6} and \ref{FIG7} for $T=T_K$ and $T=0.01T_K$,
respectively, together with its decomposition into the $f-b$ bubble
contribution (1st diagram in Fig.\ \ref{FIG5}) and the 
vertex corrections (2nd and 3rd diagrams in Fig.\ \ref{FIG5}).
Notably, even at elevated temperature, $T=T_K$, the vertex corrections
are not negligible. For $T\to 0$ the $f-b$ bubble develops an
infrared powerlaw divergence $\propto |\omega |^{\alpha _f + \alpha _b
-1}$. Since for our parameter set in the Kondo regime the exponent
$\alpha _f + \alpha _b -1 = \approx 0.056$ is rather small, and the 
pseudoparticle exponents, Eq.\ \ref{eq:FLexponents}, are obtained 
asymptotically, this singularity starts to develop for $T=0.01T_K$
merely as a discontinuity in the slope of $A_{d\sigma}(\omega)$.
Most importantly, however, the total $d-$electron spectral function
does not show any infrared singularity on the scale of $T$, and the 
width $\tilde\Gamma$ and position $\tilde\epsilon _d$ 
of the Kondo resonance are in excellent agreement with the Fermi 
liquid predictions, Eqs.\ (\ref{eq:Gtilde}), (\ref{eq:etilde}),
given the uncertainty in these quantities arising from the fact that
the Kondo resonance deviates from the Lorentzian shape for 
$\omega > T_K$ (see Appendix \ref{sec:FLrelations}). 
In contrast, the
unitarity limit, Eq.\ (\ref{eq:Ad0}), is significantly violated in 
Fig.\ \ref{FIG7}. To investigate the origin of this failure, we plot
in Fig. \ref{FIG8} the imaginary part of the interaction selfenergy
$\Sigma _{d\sigma}(\omega -i0)$. It is seen that even at the lowest
temperature the CTMA result does not develop any singularity, in
contrast to NCA. However, the position of the minimum of 
$\mbox{Im}\Sigma _{d\sigma}(\omega -i0)$ is incorrectly 
shifted to a 
negative frequency $\omega _0$ of $O(T_K)$, where 
$\mbox{Im}\Sigma _{d\sigma}(\omega _0 -i0)$ acquires a spurious
negative value. Even at the lowest $T$ considered, 
$\mbox{Im}\Sigma _{d\sigma}(\omega -i0)$ shows 
$(\omega -\omega_0)^2$ behavior for $|\omega -\omega_0| \lesssim T_K$
(Fig.\ \ref{FIG8}). Its prefactor, determined from our 
parameter set from Fig. \ \ref{FIG9} as $0.0244\;D$, is in excellent 
agreement with the exact Fermi liquid value,
$a\Gamma = 0.0239\; D$ (Eqs.\ (\ref{eq:ImSigma}),
(\ref{eq:prefactor})). The temperature dependence of the 
minimum of $\mbox{Im}\Sigma _{d\sigma}(\omega -i0)$ is analyzed 
in Fig.\ \ref{FIG10}. Again, the CTMA solution shows $T^2$ behavior 
from the lowest $T$ considered up to $T\simeq T_K$,
$\mbox{Im}\Sigma _{d\sigma}(\omega _0 -i0) = \tilde a\Gamma 
(\pi T/T_K)^2$, where the prefactor $\tilde a\Gamma = 0.013\;D$ is
of the same order of the exact value $\tilde a\Gamma$, although
roughly a factor of 2 too small. 

To summarize our analysis, the $\omega ^2$, $T^2$ behavior 
of the impurity electron selfenergy, which stems from the low-energy
excitations and is at the heart of the 
Fermi liquid theory, is even quantitatively reproduced by the CTMA
without spurious singularities. 
However, the 
location of this minimum at $\omega =0$ and the exact unitary value
$A_{d\sigma} (0)$ of the impurity spectral density are not reproduced. 
The latter is crucial to avoid a non-causal behavior of the 
correlation part of the impurity selfenergy $\Sigma _{d\sigma} (\omega )$
and, hence, for an application as an impurity solver within DMFT 
\cite{Georges.96}.  
As will be discussed below
in more detail, both of these failures can be
attributed to an incorrect treatment of non-singular potential 
scattering processes at high energies, $\omega \gg T_K$. 
We will propose a corresponding correction scheme in the next
subsection.

\begin{figure}[h!]
\centerline{\includegraphics[width=1\linewidth]{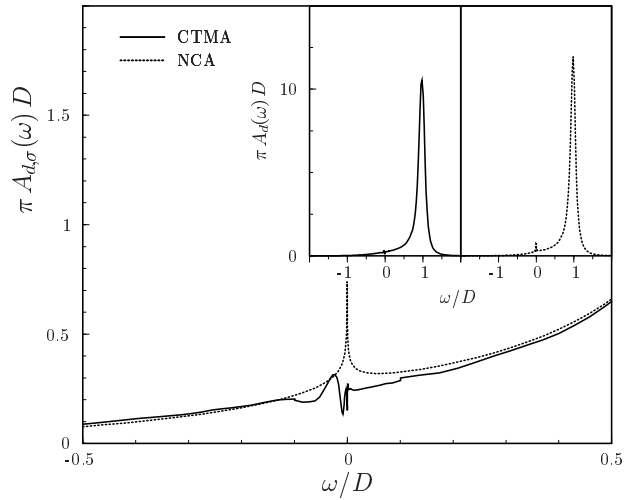}}
\caption{CTMA (solid line) and NCA (dotted line)
result for the local spectral density in the empty impurity regime,
$\epsilon _d /D = +0.81$, $\Gamma /D = 0.2$
for $T = 1.0 \times 10^{-4} D$. The resulting occupation per spin is 
$n_{d\sigma}=0.037$.  The insets show the CTMA (left panel) and the
NCA (right panel) spectral functions over 
the complete band width. } 
\label{FIG11}
\end{figure}

We conclude the present subsection by considering briefly 
another test case of Fermi liquid behavior, the empty orbital
regime. In this case the $d-$electron density of states consists of only  
one broadened, unoccupied 
single-particle resonance at a renormalized impurity level
$\epsilon _d >0$ far above the Fermi energy. In the pseudoparticle
representation, the empty-impurity case is still a non-trivial 
strongly correlated problem because of the operator constraint $Q=1$.
Fig.\ \ref{FIG11} shows the NCA and the CTMA spectral density 
for $\epsilon _d /D = +0.81$, $\Gamma /D = 0.2$. 
NCA is well known to fail badly in this case, 
producing a spurious, singular peak at $\omega =0$, which arises
from the X-ray-like divergences of the pseudoparticle Green's functions.
In the CTMA solution, the vertex corrections tend to cancel
the infrared peak of the $f-b$ bubble. Presumably,
the wiggles visible in the CTMA spectrum are due to numerical 
imprecision, but may also be due to a systematically imperfect 
cancellation of infrared 
divergent terms. In any case, the CTMA does not produce any definite
peak structure at $\omega =0$, significantly improving the 
description of the Fermi liquid behavior in the empty orbital regime.

\subsection{Effective potential scattering correction}
\begin{figure}[h!]
\centerline{\includegraphics[width=1\linewidth]{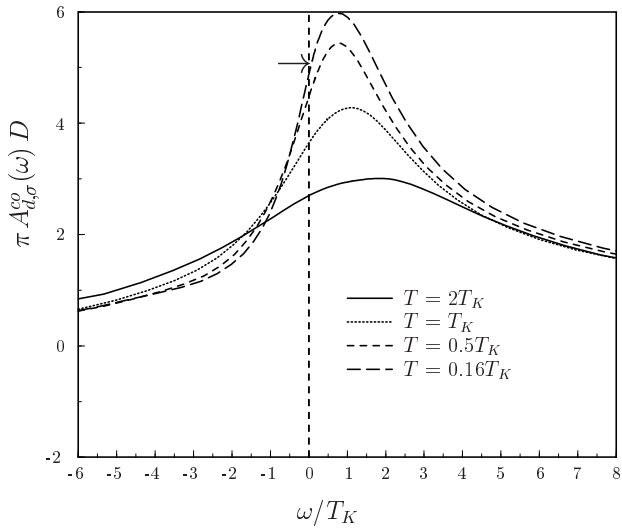}}
\caption{CTMA spectral functions with a constant potential
scattering term added to the real part of $\Sigma _{s\sigma} 
(\omega -i0)$ according to the effective potential scattering method,
see text. No low-energy singularity occurs, and the 
unitarity limit (arrow) is accurately fulfilled at the lowest temperatures. 
\label{FIG12}}
\end{figure}

\begin{figure}[h!]
\centerline{\includegraphics[width=1\linewidth]{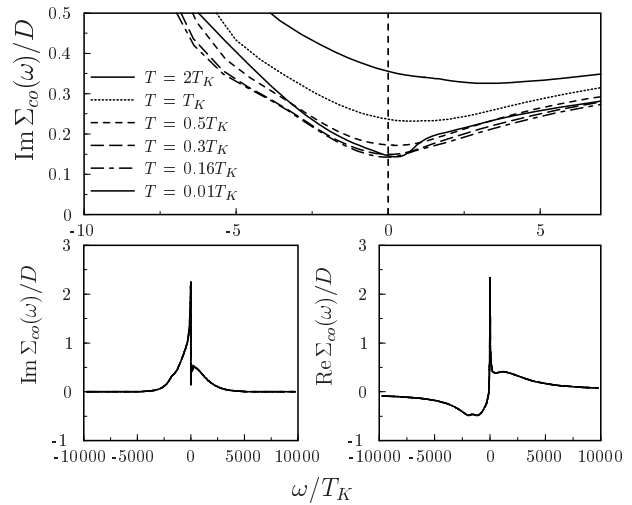}}
\caption{Real and imaginary part of the total impurity selfenergy 
$\Sigma ^{tot} _{d\sigma}(\omega -i0)$, corrected by the 
effective potential scattering method.
The curvature of $\mbox{Im}\Sigma ^{tot} _{d\sigma}(\omega -i0)$ at the Fermi
energy does not change considerably from $T=0.16T_K$ down to
$T=0.01T_K$. The small bumb in the curve for $T=0.01T_K$ at positive
$\omega$ is attributed to numerical inaccuracy.
At $\omega =0, T \to 0$, $\mbox{Im}\Sigma ^{tot} _{d\sigma}
(0 -i0)$ assumes the effective hybridization $\Gamma _{eff} \approx
0.139 D$, somewhat smaller than the exact value $\Gamma = 0.2 D$.
$\Gamma _{eff}$ is to be subtracted from 
$\mbox{Im}\Sigma ^{tot} _{d\sigma} (\omega -i0)$ in order to obtain
the interaction part of the local selfenergy.
\label{FIG13}}
\end{figure}

Within the pseudoparticle technique, it is a non-trivial task to 
separate the single-particle hybridization part $i\Gamma$, 
of the total impurity selfenergy, 
from the interaction contribution, $\Sigma _{d\sigma} (\omega )$,
since the hybridization is an interaction term in this representation,
Eq.\ (\ref{eq:hamiltonian}). Hence, while the auxiliary particle 
method is designed for a systematical treatment of the 
low-energy spin scattering processes, it is difficult to accurately
calculate the non-singular potential scattering part of the 
total impurity selfenergy 
\begin{eqnarray}
\Sigma ^{tot} _{d\sigma} (\omega ) &=&
\omega + \epsilon _F -\epsilon _d - G_{d\sigma} (\omega )^{-1}
\label{eq:Sigmatot} \\
&=& i\Gamma  + \Sigma _{d\sigma} (\omega )\ ,
\end{eqnarray}  
which involves hybridization 
processes at high energies of order $\epsilon _d$. Technically
speaking, 
the $\omega =0$ value of $Im \Sigma ^{tot} _{d\sigma} (\omega )$
is influenced by both, the real and the imaginary part of
$G_{d\sigma} (0)$, and hence by the high-energy features of the spectrum
through the Kramers-Kroenig relation.
Obtaining the precise values of the real and imaginary parts of
$\Sigma ^{tot}_{d\sigma} (0-i0)$ would therefore
require calculating the high-energy features of the local spectrum to
a precision better than $\sim T_K$. Clearly, this is a formidable 
task, both with respect to numerical precision and to diagrammatical 
systematics: Since any potential scattering term gives a 
non-singular, energy independent contribution to $\Sigma
^{tot}_{d\sigma}(0)$, it is
unlikely that a class of principal diagrams can be identified that 
reproduces the correct value. On the other hand, the class of CTMA 
diagrams does describe the correct $\omega ^2$, $T^2$ behavior,
reflecting the correct low-energy many-body dynamics. 
\newline
Based on these considerations, we propose a 
simple, phenomenological scheme to incorporate the correct potential
scattering contributions.
It amounts to adding an appropriate 
{\it frequency and temperature independent} constant $\Delta \epsilon$ to the 
{\it real part} of $\Sigma ^{tot}_{d\sigma}(\omega )$.
It has the effect of shifting the zero of the
frequency scale in {\it all} quantities, and in particular in
$\mbox{Im} \Sigma _{d\sigma} (\omega)$,  by virtue of the self-consistency. 
\cite{remark1}
Therefore, $\Delta \epsilon$ can be chosen such that at $T=0$ the minimum of 
$\mbox{Im}\Sigma _{d\sigma} (\omega -i0)$
is obtained at $\omega =0$ in accordance with Fermi liquid behavior,
Eq.\ (\ref{eq:ImSigma}). 
In the CTMA solution the position of the minimum of 
$\mbox{Im}\Sigma _{d\sigma} (\omega )$ does not significantly change with
temperature for $T \lesssim 0.2 T_K$, see Fig.\ \ref{FIG8}, as
expected from the Fermi liquid behavior, Eq.\ (\ref{eq:ImSigma}). 
Hence, we have determined $\Delta \epsilon$ from the solution at
$T=0.16T_K$ to fulfill the Fermi liquid condition above. 
The results for the impurity spectral function corrected in this way 
is displayed for various $T$ in Fig. \ref{FIG12}.
It shows accurate agreement with the unitarity limit, even though this
was not directly implied by our adjustment procedure. This can be seen
as a further indication that CTMA correctly captures the 
Fermi liquid dynamics of the problem, 
missing only part of the potential scattering contributions.  
The corresponding imaginary part of the total impurity selfenergy, 
Eq.\ (\ref{eq:Sigmatot}), is  
shown in Fig. \ref{FIG13}. Again, the Fermi liquid
behavior, Eq.\ (\ref{eq:ImSigma}), is well obeyed. The 
minimum value of $Im \Sigma ^{tot}_{d\sigma} (\omega -i0)$
approaches for $T\to 0$ 
the value $\Gamma _{eff} \approx 0.139\ D$ instead of the exact limit
$\Gamma = 0.2 \ D$. As discussed above, we attribute this to an
inaccurate treatment of single-particle hybridization processes
within CTMA. Note however, that for the DMFT algorithm
\cite{Georges.96} only the interaction part 
$\Sigma _{d\sigma} (\omega )$ of the selfenergy is important.
In the auxiliary particle method it is obtained from the
impurity Green function by the subtraction 
$\Sigma _{d\sigma} (\omega -i0) = \Sigma ^{tot}_{d\sigma} (\omega -0)
-i\Gamma _{eff} $, where  $\Sigma ^{tot}$ is given by Eq.\
(\ref{eq:Sigmatot}), and its imaginary part remains strictly non-negative.
\section{Two-channel Kondo Behavior}
\label{sec:non-FL}
To complete the discussion of dynamical quantities we calculate
the local spectral function of the two-channel
($N=2$, $M=2$) Anderson model. 
Here the low-temperature fixed point is of a distinct
non-Fermi liquid nature \cite{Nozieres.80}, 
involving a non-vanishing zero-point entropy, 
$S(0) = k_B \ln \sqrt{2}$, and a logarithmic divergence of the 
static susceptibility, $\chi (T) \propto  - \ln (T/T_K)$, 
signaling a non-degenerate ground state and over-screening of the
local spin, respectively \cite{Ludwig.91,Andrei.95}.

For the two-channel Kondo (2CK) model, the effective low-energy model 
of the two-channel SIAM, it has been shown using 
conformal field theory \cite{Affleck.93} that the local spectrum 
has a cusp at the Fermi level,  $A^{2CK} (\omega ) - A^{2CK} ( 0 ) 
\propto - |\omega| ^{1/2}$. The weight of this power law
becomes asymmetrical for $\omega >0$ and $\omega <0$, when
the particle-hole symmetry is broken, e.g., by an additional potential 
scattering term. This weight asymmetry 
is analogous to the shift of the Kondo resonance
$\tilde\epsilon _d$ in the single-channel case. Extrapolating the 
Fermi liquid results of Appendix \ref{sec:FLrelations} to the two-channel SIAM,
the weight asymmetry may be expected to be of $O(T_K / \epsilon _d)$.    
Very recently, the auxiliary particle threshold exponents for the two-channel
Anderson model have been
calculated using the Bethe ansatz.\cite{Johannesson.03} It was shown that
the exponents, like in the single-channel case, are functions of the local
valence. 

\begin{figure}[t!]
\centerline{\includegraphics[width=1\linewidth]{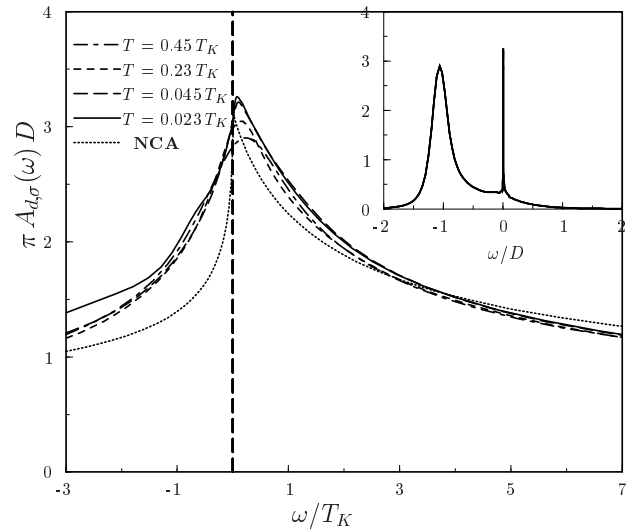}}
\caption{
CTMA results for the impurity spectral density of the
two-channel SIAM, 
with $\epsilon_d=-0.81\; D\, , \Gamma = 0.2\; D$ at various temperatures.
The NCA result is shown for comparison. The $|\omega|^{1/2}$ cusp 
develops at the Fermi level, with a weight asymmetry due to 
potential scattering. The inset shows the spectrum over a wider 
frequency range. 
\label{FIG14}}
\end{figure}

The Bethe ansatz solution shows that 
the 2CK ground state involves intricate correlations between both
conduction channels and the local spin. Thus, one would expect that
in a diagrammatic treatment three-particle correlation functions 
are needed, and that CTMA, which involves only two-particle 
$T$-matrices, is not able to capture the correct 2CK ground state. 
Indeed, in the multi-channel ($M\geq2$) case NCA as well as CTMA 
give incorrect, valence independent auxiliary particle threshold exponents 
as given in Section \ref{technique} \cite{kirchner_unpublished}. 
Surprisingly, however, 
NCA correctly reproduces qualitatively the leading low-energy
singularities of physical quantities like the susceptibility 
\cite{kirchner_unpublished} 
or the local density of states.\cite{Cox.98}
In Fig.\ \ref{FIG14} we show the CTMA solutions for the impurity spectral
function of the two-channel SIAM in comparison to the NCA result, 
both showing a $|\omega| ^{1/2}$ cusp. 
The cusp of the NCA curve has a strong weight asymmetry, which is 
presumably an overestimation, like in the Fermi liquid case (Fig.\ 
\ref{FIG6}). This is significantly improved by the CTMA solutions.
However, as mentioned above, we believe that a systematical description of the 
2CK behavior would require an extension of the CTMA to include 
three-particle correlation functions.
\section{Summary}
\label{conclusion}
To conclude, we have extended the analysis of the 
Conserving T-Matrix Approximation to thermodynamical and 
dynamical properties. 
It had been demonstrated earlier that for the single-channel Anderson impurity
model the CTMA captures the correct spin-screened Fermi liquid 
ground state on the level of the auxiliary particle dynamics,
signaled by the Fermi liquid exponents of the auxiliary particle
propagators for all fillings \cite{Kroha.97}. In the present work we
have shown that the CTMA also describes 
the Fermi liquid strong coupling behavior of physical quantities
correctly. It is, thus, the first diagrammatic method that captures
both the low-energy Fermi liquid behavior and the high-temperature
properties of quantum impurity problems of the single-channel Anderson
model type on the same footing\cite{remark2}. 
In particular the CTMA describes the  
static spin susceptibility in the Kondo regime correctly for all
temperatures. The Wilson number obtained within CTMA is in 
remarkably good agreement with the exact one.
We also showed, that the physical d-electron spectral density in the 
single-channel case ($N=2$, $M=1$) 
is able to mend most of the deficiencies of the NCA
in the Kondo and empty orbital regime of the model. Especially, no 
spurious infrared singularities occur.
The deviation of the CTMA solution from the exact unitarity limiting
value for the spectral density at the Fermi level could be traced back
to an insufficiently accurate treatment of high energy potential
scattering processes. To correct this deficiency, we have proposed 
a phenomenological method by adding an appropriate, effective potential
scattering term $\delta\epsilon$ to the real part of the impurity selfenergy, 
and defining an effective single-particle hybridization rate 
$\Gamma_{eff}$. As a result, all essential Fermi liquid properties 
are fulfilled without spurious non-causal behavior. 

Finally, we comment on the prospects for future applications of the 
CTMA. At the expense of being numerically involved, 
the CTMA combines two features which are non-trivial to fulfill by one 
single technique: 
flexibility and systematic treatment of the low-energy
excitations without {\it ad hoc} assumptions about the nature of the 
ground state. These features may make the CTMA an attractive method 
for more complicated impurity problems, such as 
(1) the selfconsistent quantum impurity problem that arises within the DMFT 
scheme \cite{Georges.96};
(2) quantum impurities with complex orbital structure; these 
arise also in cluster and cellular extensions of the DMFT 
\cite{Hettler.98,Kotliar.01,Maier.04};
(3) quantum impurity problems which may exhibit
a Fermi liquid instability.
As a diagrammatic method, the CTMA is
readily generalized for an arbitrary, energy dependent conduction electron
DOS arising from the selfconsistent DMFT scheme. In addition, 
the case of finite Coulomb repulsion $U$ must be considered, in order
to account for the upper Hubbard band and, e.g. to describe the 
metal-insulator transition in the Hubbard model near half filling. 
It requires treating the bare charge fluctuation processes involving the
empty and the doubly occupied impurity state in a symmetrical way.
On an NCA-like level (Symmetrized-U NCA, SUNCA), this has been implemented
in Ref. [\onlinecite{Haule.01}], see also [\onlinecite{Pruschke.89}],
and the corresponding Symmetrized, finite-U CTMA (SUCTMA) equations 
are reproduced in Ref. [\onlinecite{Kirchner.PhD}]. 
The SUCTMA essentially amounts to calculating, in addition to CTMA,
the ladder diagrams of heavy bosons representing the doubly occupied 
impurity state. Hence, the evaluations appear numerically manageable.
Treating a more complex impurity
orbital structure requires introducing an individual auxiliary boson
or fermion field for each charge and spin configuration of the 
impurity. Multi-orbital impurities have 
recently been treated using NCA-like approximations 
\cite{Reinert.01,Haule.04}. The fact that the number of 
impurity configurations increases roughly exponentially with the 
number of orbitals will, however, 
limit the CTMA und SUCTMA to problems with not a too
large number of local orbitals.
On the other hand, because of the systematical treatment of low-energy
excitations, the CTMA should at least be sensitive to instabilities 
of the Fermi liquid ground state due to, e.g., a quantum critical 
point.  Future developments are planned to explore the possibility of these
applications.

\begin{acknowledgments}
It is a pleasure to acknowledge stimulating  discussions with J.~Brinckmann,
T.~A.~Costi, and Q.~Si. We thank T.~A.~Costi for numerous
helpful suggestions. \mbox{S.~K.} acknowledges support by the
Deutsche Forschungsgemeinschaft and NSF Grant No. DMR-0090071. 
Additional support has been provided by SFB 195 and by SFB 608 of the DFG.
\end{acknowledgments}
\appendix
\section{CTMA Equations}
\label{sec:CTMAequations}

\begin{figure}[h!]
\centerline{\includegraphics[width=1\linewidth]{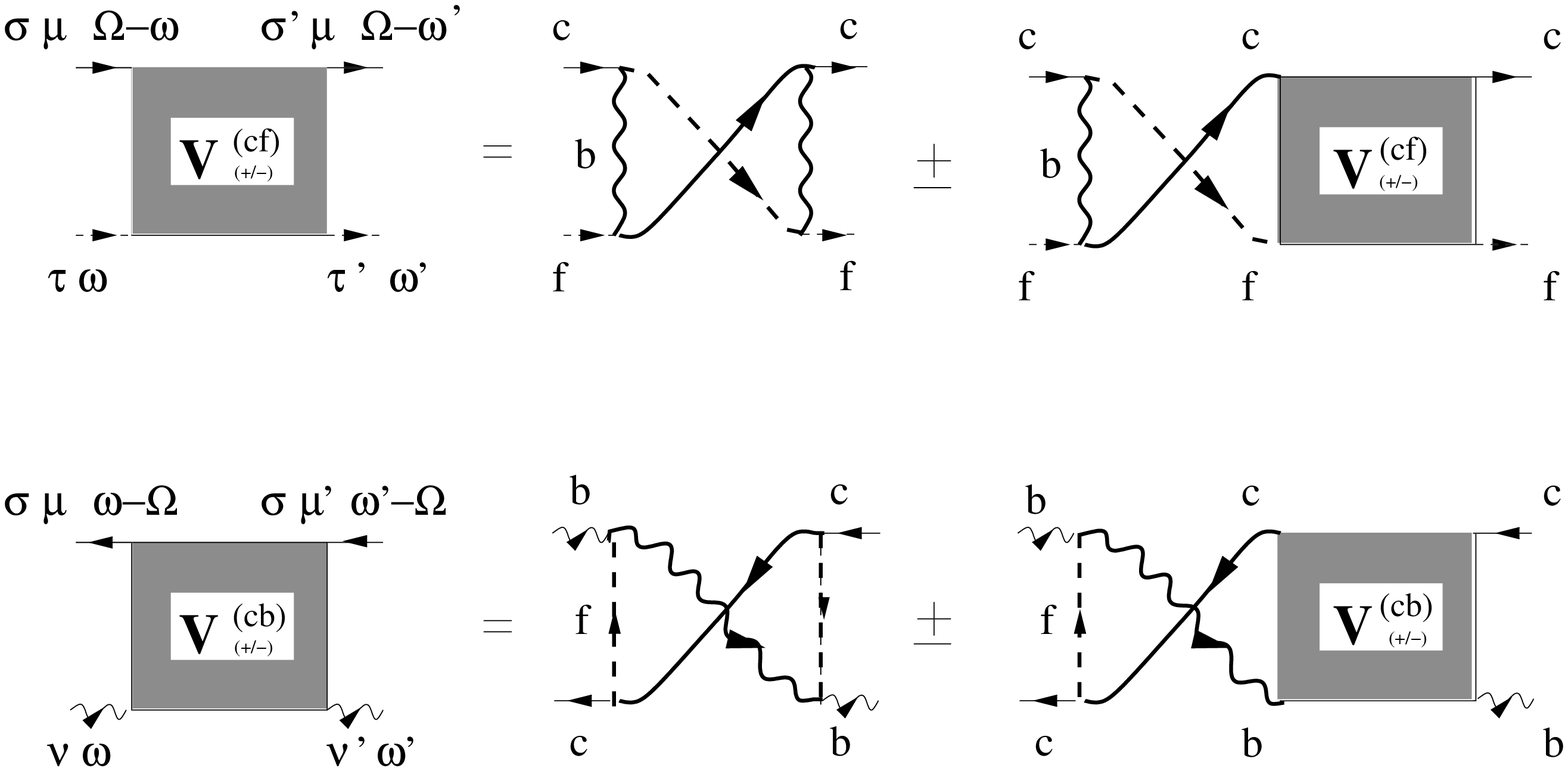}}
\vspace*{0.6cm}
\centerline{\includegraphics[width=1\linewidth]{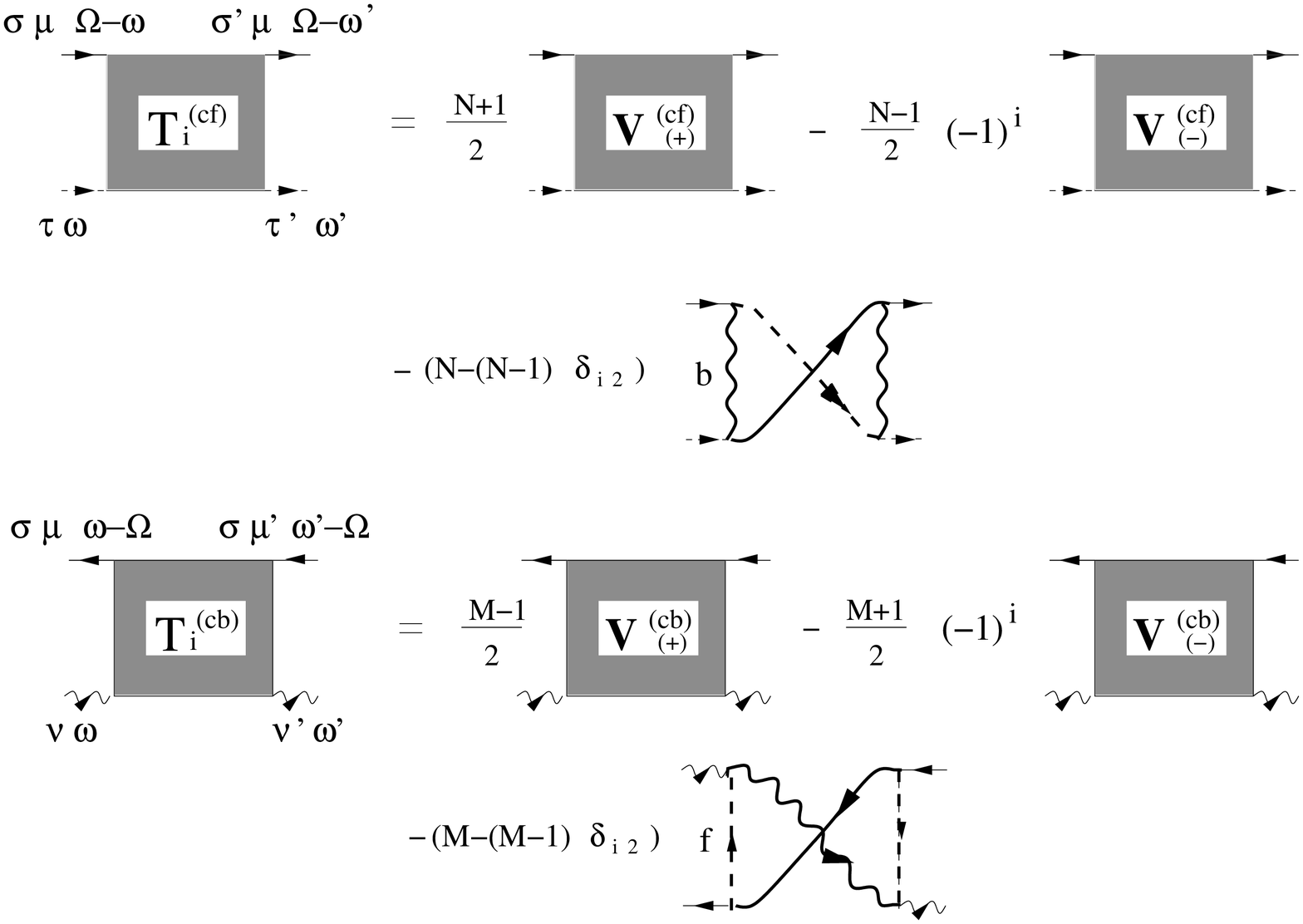}}
\vskip 4pt
\caption{Diagrammatic representation of the integral equations comprising
  the T-matrices that enter in the expressions for the
  auxiliary particle selfenergies, see the text for details.
}
\label{FIGappendix1}
\end{figure}

In the following we present a compilation of the CTMA equations which
 determine the
imaginary part of the impurity propagator $G_d$, the auxiliary particle
selfenergies $\Sigma_{f}$ and $\Sigma_{b}$ and the basic building block
of the CTMA, i.e. the analytically continued 
4-point vertices $T^{(cf)}$ and $T^{(cb)}$. Due to the fact that the
first diagram in Fig.\ref{FIG1} does not lead to a proper vertex
contribution, we start the summation with the two-rung diagram. In
order to include the proper channel and spin summations with enter
even-rung and odd-rung diagrams differently, it is necessary to solve
for the ladders with alternating signs as well.  
As already mentioned in the main text the two-rung diagrams, that is
 the inhomogeneous parts $I^{(cf)}$ and $I^{(cb)}$ of Eqs. \ref{cftmatpm} and
 \ref{cbtmatpm}
 lead to selfenergy
 contributions which are already included in the NCA. Hence, in order to avoid
 over-counting these non-skeletons have to be subtracted before
 calculating the selfenergies. The resulting equations are
 diagrammatically depicted
in Fig.\ref{FIGappendix1}.
We label the external frequencies of the 4-point vertices $V^{(cf)}$ and
$V^{(cb)}$ such, that the first (second) argument denotes the in-
(out-)going frequency of the pseudo particle propagator and the third
frequency labels the center of mass-frequency propagating through the
ladder, see Fig.~\ref{FIGappendix1}. At the Matsubara frequencies, the vertex
functions are therefore given by
\begin{widetext}
\begin{eqnarray}
V^{(cf)\,(\pm)\,\mu}_{\phantom{(f)\,(\pm)\,}\sigma ,\tau}
(i\omega _n, i\omega _n ', i\Omega _n) &=&
I^{(cf)\,\mu}_{\phantom{(f)\,}\sigma ,\tau}
(i\omega _n, i\omega _n ', i\Omega _n) 
\pm \frac{\Gamma}{\pi N(0)}\frac{1}{\beta}\sum _{\omega _n''}
G_{b\mu}(i\omega _n + i\omega _n '' - i\Omega _n )  
\label{cftmatpm}\\
&&
\times\, 
G_{f\sigma}(i\omega _n'') \ G^0_{c\mu\tau}(i\Omega _n -i\omega _n '')\ 
V^{(cf)\,(\pm)\,\mu}_{\phantom{(f)\,(\pm)\,}\tau ,\sigma}
(i\omega _n '', i\omega _n ', i\Omega _n ), \nonumber\\
I^{(cf)\,\mu}_{\phantom{(f)\,}\sigma ,\tau}
(i\omega _n, i\omega _n ', i\Omega _n) &=&
-\big(\frac{\Gamma}{\pi N(0)}\big)^2\frac{1}{\beta}\sum _{\omega _n''}
G_{b\mu}(i\omega _n + i\omega _n '' - i\Omega _n ) 
G_{f\sigma}(i\omega _n'') \ G^0_{c\mu\tau}(i\Omega _n -i\omega _n '')\ 
G_{b\mu}(i\omega _n ' + i\omega _n '' - i\Omega _n ),\nonumber
\end{eqnarray}\noindent
and
\begin{eqnarray}
V^{(cb)\,(\pm)\sigma}_{\phantom{(b)(\pm)}\mu ,\nu}
(i\omega _n, i\omega _n ', i\Omega _n ) &=&
I^{(cb)\,\sigma}_{\phantom{(b)}\mu ,\nu}
(i\omega _n, i\omega _n ', i\Omega _n ) 
\pm \frac{\Gamma}{\pi N(0)}\frac{1}{\beta}\sum _{\omega _n''}
G_{f\sigma}(i\omega _n + i\omega _n '' - i\Omega _n )  
\label{cbtmatpm}\\
&&
\times\,
G_{b\mu}(i\omega _n'') \ G^0_{c\nu\sigma}(i\omega _n ''-i\Omega _n )\ 
V^{(cb)\,(\pm)\sigma}_{\phantom{(b)(\pm)}\nu ,\mu}
(i\omega _n '', i\omega _n ', i\Omega _n ), \nonumber\\
I^{(cb)\,\sigma}_{\phantom{(b)}\mu ,\nu}
(i\omega _n, i\omega _n ', i\Omega _n ) &=&
-\big(\frac{\Gamma}{\pi N(0)} \big)^2\frac{1}{\beta}\sum _{\omega _n''}
G_{f\sigma}(i\omega _n + i\omega _n '' - i\Omega _n )  
G_{b\mu}(i\omega _n'') \ G^0_{c\nu\sigma}(i\omega _n ''-i\Omega _n )\ 
G_{f\sigma}(i\omega '_n + i\omega _n '' - i\Omega _n ). \nonumber
\end{eqnarray}\noindent
\end{widetext}
After analytical continuation 
to the real axis the T-matrices obey the
following linear Fredholm integral equations of the second kind:
\begin{widetext}
\begin{eqnarray}
V^{(cf)\,(\pm)\,\mu}_{\phantom{(f)\,(\pm)\,}\sigma ,\tau}
(\omega , \omega ' , \Omega ) &=&
I^{(cf)\,(\pm)\,\mu}_{\phantom{(f)\,(\pm)\,}\sigma ,\tau}
(\omega , \omega ' , \Omega )
\pm (-\Gamma) \int \frac{ d\varepsilon}{\pi}\;f(\varepsilon -
\Omega)\; \\
&&
\times\,
G_{b\mu}(\omega+\varepsilon-\Omega)
G_{f\sigma}(\varepsilon )  A^0_{c\mu\tau}(\Omega -\varepsilon)
V^{(cf)\,(\pm)\,\mu}_{\phantom{(f)\,(\pm)\,}\tau ,\sigma}
(\varepsilon , \omega ', \Omega  ) \nonumber\\
 I^{(cf)\,(\pm)\,\mu}_{\phantom{(f)\,(\pm)\,}\sigma ,\tau}
(\omega , \omega ' , \Omega ) &=&
\frac{\Gamma ^2}{\pi N(0)}\; \int \frac{ d\varepsilon}{\pi}\;
f(\varepsilon -\Omega) 
G_{b\mu}(\omega +\varepsilon -\Omega)
G_{f\sigma}(\varepsilon )  A^0_{c\mu\tau}(\Omega -\varepsilon)
G_{b\mu}(\omega '  +\varepsilon -\Omega)  \nonumber
\label{tcfcontinued}
\end{eqnarray}
and
\begin{eqnarray}
V^{(cb)\,(\pm)\sigma}_{\phantom{(b)(\pm)}\mu ,\nu}
(\omega , \omega ', \Omega ) &=&
I^{(cb)\,(\pm)\sigma}_{\phantom{(b)(\pm)}\mu ,\nu}
(\omega , \omega ', \Omega )
\pm (+\Gamma) \int \frac{ d\varepsilon}{\pi}\;f(\varepsilon -\Omega )\;\\
&&
\times\,
G_{f\sigma}(\omega +\varepsilon -\Omega )
G_{b\mu} (\varepsilon )
A^0_{c\nu\sigma}(\varepsilon -\Omega )\
V^{(cb)\,(\pm)\sigma}_{\phantom{(b)(\pm)}\nu ,\mu}
(\varepsilon , \omega ', \Omega ) \nonumber\\
 I^{(cb)\,(\pm)\sigma}_{\phantom{(b)(\pm)}\mu ,\nu}
(\omega , \omega ', \Omega ) &=&
-\frac{\Gamma ^2}{\pi N(0)}\; \int \frac{ d\varepsilon}{\pi}\;
f(\varepsilon -\Omega)  
G_{f\sigma}(\omega + \varepsilon -\Omega )
G_{b\mu}(\varepsilon ) A^0_{c\nu\sigma}(\varepsilon -\Omega)\
G_{f\sigma}(\omega '+ \varepsilon -\Omega ). \nonumber
\label{tcbcontinued}
\end{eqnarray}
\end{widetext}

\begin{figure}[h!]
\vskip 8pt
\centerline{\includegraphics[width=1\linewidth]{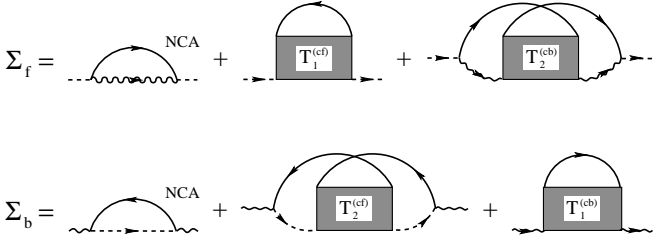}}
\vskip 4pt
\caption{Auxiliary particle selfenergies $\Sigma_{f}$ and $\Sigma_b$:
  The first diagram is the NCA contribution, where however fully
  CTMA-renormalized Green functions have to be used. The second and
  third diagram constitute the vertex corrections from $T^{(fc)}$ and $T^{(bc)}$.
}
\label{FIGappendix2}
\end{figure}
In order to simplify the expressions for the selfenergies
$\Sigma_{f,\sigma}$
 and $\Sigma_b$ it proves useful to introduce
\begin{eqnarray*}
T^{(cf)}_1 &=& \frac{N+1}{2}\;V^{(cf)\,(+)} + \frac{N-1}{2}\; V^{(cf)\,(-)}
               -N\, I^{(cf)},\label{tcf1}\\
T^{(cf)}_2 &=& \frac{N+1}{2}\;V^{(cf)\,(+)} - \frac{N-1}{2}\; V^{(cf)\,(-)}
               - I^{(cf)}, \label{tcf2}
\end{eqnarray*}
and
\begin{eqnarray*}
T^{(cb)}_1 &=& \frac{M-1}{2}\; V^{(cb)\,(+)} + \frac{M+1}{2}\; 
               V^{(cb)\,(-)}-M\, I^{(cb)},\label{tcb1}\\
T^{(cb)}_2 &=& \frac{M-1}{2}\; V^{(cb)\,(+)} - \frac{M+1}{2}\; 
               V^{(cb)\,(-)}-I^{(cb)}\; .\label{tcb2}
\end{eqnarray*}
Then we obtain for the analytically continued advanced ($i\omega
\rightarrow \omega-i0 \equiv \omega$) selfenergies (Fig. \ref{FIGappendix2}):
\begin{eqnarray}
\Sigma_{f\sigma}^{}(\omega)&=& \Sigma_{f\sigma}^{\mbox{\tiny
    (NCA)}}(\omega)+\Sigma_{f\sigma}^{(cf)}(\omega)+\Sigma _{f\sigma}^{(cb)}(\omega)\\
\Sigma_{b\mu}^{}(\omega)&=& \Sigma_{b\mu}^{\mbox{\tiny
    (NCA)}}(\omega)+\Sigma_{b\mu}^{(cf)}(\omega)+\Sigma _{b\mu}^{(cb)}(\omega)
\end{eqnarray}
with
\begin{eqnarray*}
\Sigma_{f\sigma}^{\mbox{\tiny (NCA)}}(\omega)&=&M\Gamma\sum _{\mu}\int
              \frac{{d}\varepsilon}{\pi}\,
               f(-\varepsilon ),
              A_{c\mu\sigma}^0(\varepsilon)G_{b\mu}(\omega -\varepsilon)
             \\ \label{sigfNCA}
\Sigma_{b\mu}^{\mbox{\tiny (NCA)}}(\omega)&=&N\Gamma\sum _{\sigma}\int
              \frac{{d}\varepsilon}{\pi}\,
              f(\varepsilon )A_{c\mu\sigma}^0(\varepsilon)
              G_{f\sigma}(\omega +\varepsilon),
              \label{sigbNCA}
\end{eqnarray*}

\begin{eqnarray*}
\Sigma _{f\sigma}^{(cf)}(\omega ) &=& M
\int \frac{ d\varepsilon}{\pi}\;f(\varepsilon -\omega)\;
A^0_{c}(\varepsilon -\omega)\;  \\ \nonumber
&&
\times\,
 \pi{ N}(0)
T^{(cf)}_1 (\omega ,\omega ,\varepsilon ),\label{sigffcontinued}\\ \nonumber
\Sigma _{f\sigma}^{(cb)}(\omega ) &=& 
-M\; \Gamma \int \frac{ d\varepsilon}{\pi}\int \frac{ d\varepsilon '}{\pi}
\;f(\varepsilon -\omega)\;f(\varepsilon ' -\omega)\;
\label{sigfbcontinued} \\ \nonumber
&&
\times\,
A^0_{c}(\omega -\varepsilon )
G _{b}(\varepsilon )
\pi{ N}(0) T^{(cb)}_2 (\varepsilon ,\varepsilon ' ,
\varepsilon +\varepsilon ' -\omega ) \\ \nonumber
&&
\times\,
A^0_{c}(\omega -\varepsilon ') G _{b}(\varepsilon ' ),\nonumber
\end{eqnarray*}
and 
\begin{eqnarray*}
\Sigma _{b\mu}^{(cf)}(\omega ) &=& 
-N\; \Gamma \int \frac{ d\varepsilon}{\pi}\int \frac{ d\varepsilon '}{\pi}
\;f(\varepsilon -\omega)\;f(\varepsilon '-\omega )\;
\label{sigbfcontinued}\\ 
&&
\times\,
A^0_{c}(\varepsilon -\omega )
G _{f}(\varepsilon )\;
\pi{ N}(0)T^{(cf)}_2 (\varepsilon ,\varepsilon ' ,
\varepsilon +\varepsilon ' -\omega )  \\ 
&&
\times\,
A^0_{c}(\varepsilon ' -\omega)
G _{f}(\varepsilon ' ), \\
\Sigma _{b\mu}^{(cb)}(\omega ) &=& -N
\int \frac{ d\varepsilon}{\pi}\;f(\varepsilon -\omega)\;
A^0_{c}(\omega -\varepsilon ) \\ 
&&
\times\,
\pi{ N}(0) T^{(cb)}_1 (\omega ,\omega ,\varepsilon ) ,
\label{sigbbcontinued}
\end{eqnarray*}
\begin{figure}[ht!]
\vskip 8pt
\centerline{\includegraphics[width=1\linewidth]{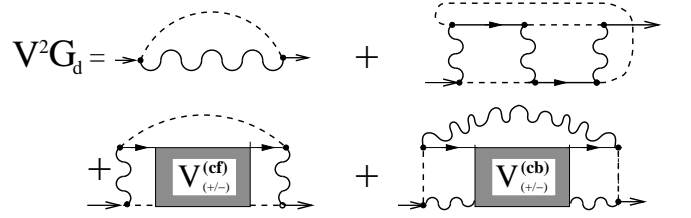}}
\caption{
  Diagrammatic representation of the equation for the local
  d-electron Green function. The first diagram is the NCA
  contribution. As described in the text, because of the
  analytical structure of the Adsigma it is necessary to
  multiply with a rung at each end of the vertex functions
 $V^{(cf)}_{(\pm)}$, $V^{(cb)}_{(\pm)}$,
  as shown in the last two diagrams on the right-hand side.
  Hence the sum of these diagrams
  contains at lease 4 rungs, i.e. the three-rung term must be
  added separately (second diagram on the right-hand side.)
}
\label{FIGappendix3}
\end{figure}
where the vertex functions follow from $T^{(cf)}_{1/2}$ and $T^{(cb)}_{1/2}$ after performing the frequency summations and 
analytical continuation of the external frequencies. In the contour integrals around conduction line cuts we have used
$A_{c}^0(\omega )=(1\pi)\, {\rm Im}G_{c\mu\sigma}^0(\omega-i0)/N(0)$. 
\newline
Once the pseudo particle Green functions have been self-consistently
calculated, physical properties can be determined. 
In order to construct the local Green function from
the basic building block of our theory, the 4-point vertices $T^{(cf)}$ and
$T^{(cb)}$, we note that the Bethe-Salpeter equations above yield either
fully advanced or fully retarded T-matrices,
\begin{eqnarray}
&&T^{RRR}_{}(\omega,\omega^{\prime},\Omega)\equiv  \\ 
&&~~~~~ T(i\omega \rightarrow\omega + i0,i\omega^{\prime} \rightarrow \omega^{\prime} + i0,i\Omega \rightarrow
\Omega + i0)  \nonumber
\end{eqnarray}
or
\begin{eqnarray}
&&T^{AAA}_{}(\omega,\omega^{\prime},\Omega)\equiv   \\
&& ~~~~~T(i\omega \rightarrow
\omega -i0,i\omega^{\prime} \rightarrow \omega^{\prime}- i0,i\Omega \rightarrow
\Omega - i0)  \nonumber
\end{eqnarray}

The local spectral density  however requires the determination of
$T^{RRA}(\omega,\omega^{\prime},\Omega)$.
Therefore, its calculation is more involved. We construct
$T^{RRA}(\omega,\omega^{\prime},\Omega)$ by adding two rungs to each of
the diagrams in  $T^{(cf)}$ and $T^{(cb)}$. This then turns for example the two-rung
diagram into a four-rung ladder. In order to include all contributing
diagrams, it is then necessary to employ the T-matrices $V^{(cf)}_{(\pm)}$
and $V^{(cb)}_{(\pm)}$ instead of  $T^{(cf)}$ and $T^{(cb)}$ and, in addition,
the three-rung diagram has to be added, see Fig. \ref{FIGappendix3}. 
The equation for the advanced spectral function after projection onto
the physical subspace then follows as 
\begin{widetext}
  \begin{eqnarray}
A_{d,\sigma}(\omega)&=& A_{d,\sigma}^{\mbox{\tiny (NCA)}}(\omega)\,+\,\frac{\mathcal{N}(0)}{\pi^2 \Gamma}\Bigg[ \,
\int^{\infty}_{\infty}d\tilde{\epsilon} \frac{e^{-\beta \tilde{\epsilon}}}{f(\omega)}
\int^{\infty}_{\infty}d\epsilon_1 \, \int^{\infty}_{\infty}d\epsilon_2 \, f(\epsilon_1-\tilde{\epsilon})\, f(\epsilon_2-\tilde{\epsilon})
A^{0}_{c \sigma\mu}(\tilde{\epsilon}-\epsilon_1)\,A^{0}_{c \sigma\mu}(\tilde{\epsilon}-\epsilon_2) \nonumber \\
&& \times\, \mathop{\rm Im} \big\{ G_{f\sigma}(\epsilon_1)\,G_{f\sigma}(\epsilon_2)\,G_b(\epsilon_1+\epsilon_2-\tilde{\epsilon})\big\}\cdot
  \mathop{\rm Im} \big\{ G_{f\sigma}(\tilde{\epsilon}-\omega)\,G_{b}(\epsilon_1-\omega)\,G_b(\epsilon_2-\omega)\big\} \nonumber \\[1.5ex]
&+&\, \int^{\infty}_{\infty}d\tilde{\epsilon} \frac{e^{-\beta \tilde{\epsilon}}}{f(\omega)}
\int^{\infty}_{\infty}d\epsilon_1 \, \int^{\infty}_{\infty}d\epsilon_2 \, f(\epsilon_1-\tilde{\epsilon})\, f(\epsilon_2-\tilde{\epsilon})
A^{0}_{c \sigma\mu}(\tilde{\epsilon}-\epsilon_1)\,A^{0}_{c \sigma\mu}(\tilde{\epsilon}-\epsilon_2) \nonumber \\
&&\times\,  \mathop{\rm Im} \big\{ G_{f\sigma}(\epsilon_1)\,G_{f\sigma}(\epsilon_2)\,
\big[\frac{N-1}{2}\cdot V^{cf}_{+,\sigma,\sigma^{\prime}}(\epsilon_1,\epsilon_2,\tilde{\epsilon})+\frac{N+1}{2}\cdot
 V^{cf}_{-,\sigma,\sigma^{\prime}}(\epsilon_1,\epsilon_2,\tilde{\epsilon})
\big]\big\} \nonumber \\
&& \times\, \mathop{\rm Im} \big\{ G_{f\sigma}(\tilde{\epsilon}-\omega)\,G_{b\mu}(\epsilon_1-\omega)\,G_{b\mu}(\epsilon_2-\omega)\big\}  \\[1.5ex]
&+&\, \int^{\infty}_{\infty}d\tilde{\epsilon} \frac{e^{-\beta \tilde{\epsilon}}}{f(-\omega)}
\int^{\infty}_{\infty}d\epsilon_1 \, \int^{\infty}_{\infty}d\epsilon_2 \, f(\epsilon_1-\tilde{\epsilon})\, f(\epsilon_2-\tilde{\epsilon})
A^{0}_{c \sigma\mu}(\epsilon_1-\tilde{\epsilon})\,A^{0}_{c \sigma\mu}(\epsilon_2-\tilde{\epsilon}) \nonumber \\
&&\times\,  \mathop{\rm Im} \big\{ G_{b\mu}(\epsilon_1)\,G_{b\mu}(\epsilon_2)\,
\cdot[\frac{M-1}{2}\cdot V^{cb}_{+,\sigma,\sigma}(\epsilon_1,\epsilon_2,\tilde{\epsilon})+\frac{M+1}{2}\cdot V^{cb}_{-,\sigma,\sigma}(\epsilon_1,\epsilon_2,\tilde{\epsilon}) ]\big\} \nonumber \\
&&\times\,  \mathop{\rm Im} \big\{ G_{b\mu}(\tilde{\epsilon}+\omega)\,G_{f\sigma}(\epsilon_1+\omega)\,G_{f\sigma}(\epsilon_2+\omega)\big\}
\Bigg], \nonumber
\end{eqnarray}
with
\begin{equation}
A_{d,\sigma}^{\mbox{\tiny (NCA)}}(\omega)
         = \int  {d}\varepsilon\,  {\rm e}^{-\beta\varepsilon}
         [ A_{f\sigma}(\omega +\varepsilon )A_{b\mu}(\varepsilon )
-A_{f\sigma}(\varepsilon )
                   A_{b\mu}(\varepsilon -\omega) ].\label{gdNCA}
\end{equation}
\end{widetext}
\section{Fermi Liquid Relations}
\label{sec:FLrelations}
In this appendix we compile some exact Fermi liquid properties for
the d-electron Green's function ($N=2$, $M=1$),
\begin{equation}
G _{d\sigma} (\omega \pm i0) = 
\frac{1}{\omega +\epsilon _F -\epsilon _d \pm i\Gamma - \Sigma _{d\sigma}
(\omega \pm i0) }  \ .
\label{eq:AppendGd}
\end{equation}
The interaction contribution $\Sigma _{d\sigma} (\omega \pm i0 )$
to the selfenergy obeys the Fermi liquid relations
\begin{eqnarray}
\mbox{Im} \Sigma _{d\sigma } (\omega \pm i0) = 
a \,\Gamma\, \frac{\omega ^2 + (\pi T)^2}{T_K^2} \ &,& \omega , T \lesssim T_K 
\label{eq:AppendImSigma}\\
\int _{-\infty} ^{0} d\omega 
\frac{\partial \Sigma _{d\sigma}}{\partial \omega} G_{d\sigma} (\omega ) =0
\ &,& T=0 
\label{eq:AppendLuttinger}
\end{eqnarray}
These imply the Friedel sum rule, $n _{d\sigma} (0) = \delta _{\sigma}$,
with $n_{d\sigma} (0)$ the impurity occupation at $T=0$ and $\delta _{\sigma}$
the scattering phase at the Fermi energy $\epsilon _F$,
\begin{eqnarray}
\cot \delta _{\sigma} = \frac{\mbox{Re}G_{d\sigma}(0)}
{\mbox{Im}G_{d\sigma}(0+i0)}\ .
\label{eq:AppendPhase}
\end{eqnarray}
The impurity spectral density at the Fermi energy $\omega =0$, $T=0$
follows from Eqs.\ (\ref{eq:AppendGd}), (\ref{eq:AppendImSigma}), 
(\ref{eq:AppendPhase}) as
\begin{equation}
A_{d\sigma} = \frac{1}{\pi} \mbox{Im} G_{d\sigma} (0 -0) =
\frac{\sin ^2 (\pi n_{d\sigma})}{\pi \Gamma}. 
\label{eq:AppendAd0}
\end{equation}
The Fermi liquid relations also determine the width $\tilde\Gamma$, 
position $\tilde\epsilon _d$, and spectral weight $z$ of the
Kondo resonance. Using the quasiparticle weight 
$z=(1-\partial \Sigma _{d\sigma} /\partial \omega |_{\omega =0} )^{-1}$,
the Green's function can be written for $|\omega|,\ T \lesssim T_K$,
\begin{equation}
G _{d\sigma} (\omega \pm i0) = 
\frac{z}{\omega -\tilde \epsilon _d \pm i\tilde \Gamma }  \ ,
\label{eq:AppendGqp}
\end{equation}
with $\tilde\epsilon _{d} = z [\epsilon _F +\epsilon _d + 
\mbox{Re} \Sigma _{d\sigma} (0) ]$, $\tilde\Gamma = z \Gamma$.
z can be expressed in terms of the strong coupling parameters of
the Anderson model by equating the $T=0$ spin susceptibility,
Eq.\ (\ref{eq:chi0}), with the Pauli susceptibility of a Fermi liquid
of quasiparticles with weight $z$, 
$\chi(0) = (g\mu _B)^2 A_{d\sigma} /z$,
\begin{equation}
z=\frac{4 T_K^{\star} \sin^2(\pi n_{d\sigma}) }{\pi \Gamma W },
\end{equation}
and hence,
\begin{equation}
\tilde\Gamma = \frac{4}{\pi W} \sin ^2 (\pi n_{d\sigma}) T_K^{\star}\ .
\label{eq:AppendGammatilde}
\end{equation}
Combining Eqs.\ (\ref{eq:AppendPhase}), (\ref{eq:AppendGqp}),
(\ref{eq:AppendGammatilde})
the position of the Kondo resonance relative to $\epsilon _F$ 
is given by
\begin{equation}
\tilde\epsilon _d = \frac{2}{\pi W} \sin  (2 \pi n_{d\sigma}) T_K^{\star}\ . 
\end{equation}
The exact prefactor $a$ of the quadratic behavior of 
$\mbox{Im} \Sigma _{d\sigma} (\omega )$ has been calculated for the
Anderson model using direct perturbation theory  to infinite order 
in the on-site repulsion $U$ \cite{Hewson,Yamada.76,Yosida.66}.
One obtains at $T=0$, 
$\partial ^2 \mbox{Im} \Sigma _{d\sigma } (\omega - i0) / \partial \omega ^2 
= \pi |\gamma (0,0)|^2 [A_{d\sigma}(0)]^3$, where $\gamma (0,0)$ is the 
full, local 2-electron vertex function at the Fermi energy and at $T=0$.
Using a Ward identity \cite{Hewson}, it can be related to the 
quasiparticle density of states $A_{d\sigma}(0)/z$ and expressed as
$\gamma (0,0) = (R-1)/[zA_{d\sigma}(0)]$, where $R=2$ is the 
Wilson ratio. Hence, taking into account Eq.\ (\ref{eq:TKW}), 
we obtain for the prefactor in Eq.\ (\ref{eq:AppendImSigma}),
\begin{equation}
a = \frac{\pi^4 W^2}{8 \mbox{e}^{3/2+2C}} \; 
\frac{(R-1)^2}{\sin^2(\pi n_{d\sigma})}\; 
\frac{|\epsilon _d|}{D} \ .
\end{equation}


\begin{thebibliography}{62}
\bibitem[{rem()}]{remark1}
\bibinfo{note}{Note that $\Delta\epsilon$ is analogous to a chemical
  potential. Since the shift $\Delta \epsilon \approx T_K \ll \epsilon _d,\
  \Gamma$, and the impurity occupation $n_{d\sigma}$ is predominantly
  determined by the single-particle resonance at $\omega \approx \epsilon _d$,
  the resulting change of $n_{d\sigma}$ is of $O(\; T_K /[\Gamma (1+(\epsilon
  _d /\Gamma )^2)]\; )$, i.e. negligibly small.}
\bibitem[{rem()}]{remark2}
\bibinfo{note}{By contrast, the local moment approach (LMA) of
Ref. [\onlinecite{Logan.98,Logan.00,Logan.02,Glossop.02,Glossop.03a,
Glossop.03b}]
is based on an uncontrolled approximation: the superposition of broken
symmetry states and
an ad hoc effective interaction determined by local constraints. The LMA
provides a reasonable, although not in every respect correct interpolation
between the high and low temperature regimes in the single channel case. It
is inapplicable to the multichannel case.}
\end{thebibliography}
\end{document}